\documentclass[english]{article}
\usepackage[T1]{fontenc}
\usepackage[latin9]{inputenc}
\usepackage{geometry}
\usepackage{amsmath}
\usepackage{amssymb}
\usepackage{setspace}
\usepackage{esint}
\usepackage[all]{xy}
\makeatletter
\newcommand{\lyxaddress}[1]{
\par {\raggedright #1
\vspace{1.4em}
\noindent\par}
}

\usepackage{babel}
\usepackage{babel}
\newcommand{\xyR}[1]{\xydef@\xymatrixrowsep@{#1}}
\newcommand{\xyC}[1]{\xydef@\xymatrixcolsep@{#1}}

\usepackage{babel}

\makeatother

\usepackage{babel}
\begin{document}
\begin{center}
{\LARGE{Analytical continuum mechanics à la Hamilton-Piola:}} 
\par\end{center}

\begin{center}
{\LARGE{least action principle for second gradient continua and capillary
fluids}} 
\par\end{center}

\begin{center}
By N. Auffray$^{\text{a}}$, F. dell'Isola$^{\text{b}}$, V. Eremeyev$^{\text{c}}$,
A. Madeo$^{\text{d}}$ and G. Rosi$^{\text{e}}$ 
\par\end{center}

\lyxaddress{{\scriptsize{$^{\text{a}}$Université Paris-Est, Laboratoire Modélisation
et Simulation Multi Echelle, MSME UMR 8208 CNRS, 5 bd Descartes, 77454
Marne-la-Vallée, France}}\\
{\scriptsize{$^{\text{b}}$Dipartimento di Ingegneria Strutturale
e Geotecnica, Università di Roma La Sapienza, Via Eudossiana 18, 00184,
Roma, Italy }}\\
{\scriptsize{$^{\text{c}}$Institut für Mechanik, Otto-von-Guericke-Universität
Magdeburg, 39106 Magdeburg, Germany, and}}\\
{\scriptsize{South Scientific Center of RASci \& South Federal University,
Rostov on Don, Russia}}\\
{\scriptsize{$^{\text{d}}$Laboratoire de Génie Civil et Ingénierie
Environnementale, Université de Lyon--INSA, Bâtiment Coulomb, 69621
Villeurbanne Cedex, France}}\\
{\scriptsize{$^{\text{e}}$International Center MeMOCS ``Mathematics
and Mechanics of Complex System'', Università degli studi dell'Aquila,
Palazzo Caetani, Via San Pasquale snc, Cisterna di Latina, Italy}}}
\begin{quotation}
\emph{``On ne trouvera point de Figures dans cet Ouvrage. Les méthodes
que j'y expose ne demandent ni constructions, ni raisonnemens géométriques
ou méchaniques, mais seulement des opérations algébriques, assujetties
à une marche réguliere et uniforme. Ceux qui aiment l'Analyse, verront
avec plaisir la Méchanique en divenir une nouvelle branche, et me
sauront gré d'en avoir étendu ansi le domaine.''}

\begin{flushright}
\textit{\emph{From the}}\textit{ Avertissement of the Méchanique Analitique
}\textit{\emph{by}}\textit{ Lagrange \cite{Lagrange1788}} 
\par\end{flushright}
\end{quotation}

\section{Abstract}

In this paper a \emph{stationary action principle} is proved to hold
for capillary fluids, i.e. fluids for which the deformation energy
has the form suggested, starting from molecular arguments, for instance
by Cahn and Hilliard \cite{Cahn-Hilliard1958,Cahn-Hilliard1959}.
We remark that these fluids are sometimes also called Korteweg-de
Vries or Cahn-Allen fluids. In general continua whose deformation
energy depends on the second gradient of placement are called \textit{second
gradient} (or \textit{Piola-Toupin} or \textit{Mindlin} or \textit{Green-Rivlin}
or \textit{Germain} or \textit{second grade)} continua. In the present
paper, a material description for second gradient continua is formulated.
A Lagrangian action is introduced in both the material and spatial
descriptions and the corresponding Euler-Lagrange equations and boundary
conditions are found. These conditions are formulated in terms of
an objective deformation energy volume density in two cases: when
this energy is assumed to depend on either $C$ and $\nabla C$ or
on $C^{-1}$ and $\nabla C^{-1},$ where $C$ is the Cauchy-Green
deformation tensor. When particularized to energies which characterize
fluid materials, the capillary fluid evolution conditions (see e.g.
Casal \cite{CASAL1972} or Seppecher \cite{SEPPECHER1988,SEPPECHER1996}
for an alternative deduction based on thermodynamic arguments) are
recovered. A version of Bernoulli's law valid for capillary fluids
is found and, in Appendix B, useful kinematic formulas for the present
variational formulation are proposed. Historical comments about Gabrio
Piola's contribution to analytical continuum mechanics are also presented.
In this context the reader is also referred to Capecchi and Ruta \cite{CapRut07}.

\part{Introduction}

Since its first formulation, which can be attributed to D'Alembert
and Lagrange, continuum mechanics has been founded on the \emph{principle
of virtual work}. Moreover, since the early modern%
\footnote{It is well known that Archimedes could formulate a precise theory
of the equilibrium of fluids (see e.g. Rorres \cite{Rorres}) and
there are serious hints that a form of Bernouilli law for fluid flow
was known to Hellenistic scientists (see e.g. Vailati \cite{VAILATI1897}
or Russo \cite{Russo}).%
} studies on the equilibrium and motion of fluids, the concept of a
continuous body was considered suitable to model macroscopic mechanical
phenomena. On the other hand, Poisson \cite{Poisson1,Poisson2,poisson3}
preferred, instead, a treatment based on an atomistic or molecular
point of view. As the actual configuration of a continuous system
is characterized by a placement function (see Piola \cite{Piola1,Piola2,Piola4}
for one of the first precise presentation of the analytical concepts
involved in this statement) one can clearly see the main mathematical
difference between discrete and continuous models: the configuration
space is finite dimensional in the first case and infinite dimensional
in the second one. Indeed a configuration is characterized as a $n-tuple$
of real variables (Lagrange parameters) when introducing discrete
models or as a set of fields, defined on suitably fixed domains, when
introducing continuous models. Of course the comparison of the two
modeling approaches has to be based on the different relevant physical
aspects of the considered phenomena. The reader is referred to the
vivid discussion of this point already presented by Piola (see infra
in the following subsections and in particular his discussion about
reality as perceived by the \textit{animaletti infusorj} (i.e. micro-organisms)).
It was clear already to Euler, D'Alembert and Lagrange \cite{Lagrange1788}
that, in order to formulate an effective model to describe a large
class of physical phenomena occurring in deformable bodies, it can
be more convenient to introduce a set of space-time partial differential
equations for a small number of fields (i.e. functions defined in
suitably regular subsets of $\mathbb{R}^{3})$ instead of a set of
ordinary differential equations in which the set of unknown functions
outnumbers any imaginable cardinality.

The fundamental conceptual tool used in continuous models is the definition
of the so-called Lagrangian configuration, in which any material particle
of the considered continuous body is labeled by three real variables,
the material (or Lagrangian) coordinates of the considered particle.
As a consequence the motion of a continuous system is characterized
by the time dependence of the chosen set of fields. For both discrete
and continuous models the obvious problem arises, once the space of
configurations is fixed and the set of admissible motions chosen,
how to determine the equations of motion ? In other words: how to
model the external interactions between the external world, the considered
body and the internal interactions of the body in order to get some
evolution equations which, once solved, supply a reliable prediction
of the body behavior ?

There are different postulation schemes which, during the centuries,
have been proposed to that aim. All of these schemes have their merits
and their defects: with a somehow inappropriate simplification we
have classified them into two subgroups (see a subsection infra) gathered
under the collective names \emph{analytical continuum mechanics} and
\emph{continuum thermodynamics}. It has to be emphasized that some
remarkable results were obtained by combining the two approaches:
in the present context one has to cite the works by Seppecher \cite{SEPPECHER1988,SEPPECHER1996}.
In these papers the author obtained the evolution equations for capillary
fluids by combining the principle of virtual works in the Eulerian
description with the first principle of thermodynamics (also in the
case of isothermal motions). This shows that it can be sometimes useful
to use an heuristic procedure in which the principle of virtual powers
is reinforced by additionally requiring also the validity of the balance
of mechanical energy. Also interesting in this context are the results
presented in Casal \cite{CASAL1972}, Gavrilyuk and Gouin \cite{GavrilyukGouin1999}.

In the opinion of the present authors the methods of analytical continuum
mechanics are the most effective ones (see also \cite{Maugin2011}),
at least when formulating models for mechanical phenomena involving
multiple time and length scales. The reader is invited to consider,
with respect to this particular class of phenomena, the difficulties
which are to be confronted when using continuum thermodynamics, for
instance, to describe interfacial phenomena in phase transition (see
e.g. dell'Isola and Romano \cite{dellisolaromano1,dellisolaromano2,dell'isolaromano3}
and dell'Isola and Kosinski \cite{dellisolakosinski1993}, or in poroelasticity
see e.g. dell'Isola and Hutter \cite{dellisolahutter(1998)}). These
difficulties are elegantly overcome when accepting to use as a fundamental
tool the principle of virtual work (as done in Casal and Gouin \cite{Casal-Gouin1985},
Seppecher \cite{SEPPECHER1993} and dell'Isola et al. \cite{dellisolaseppechermadeo2012}).
Related phenomena occur in the flow of bubbles surrounded by their
liquid phase: it could be interesting to apply the homogenization
techniques presented in Boutin and Auriault \cite{Boutin1} to the
equations for capillary fluids presented here.

\subsection{Deduction of the evolution equations for capillary fluids and second
gradient solids by the principle of least action}

In the present paper we show that it is possible to deduce from the
principle of least action, and without any further assumption, the
whole set of evolution equations (i.e. bulk equations and boundary
conditions) for capillary fluids both in the Lagrangian and Eulerian
descriptions. These equations are the Euler-Lagrange conditions corresponding
to a precisely specified action functional. The obtained variational
principle will be useful at least when formulating numerical schemes
for studying a large class of flows of capillary fluids. Also a form
of Bernoulli's law, valid for capillary fluids, is here observed to
hold. Moreover we find the complete Lagrangian form of the evolution
equations for second gradient solids when the deformation energy is
assumed to depend on the deformation measure $C:=F^{T}F$ (where $F$
is the placement gradient with respect to Lagrangian referential coordinates)
or, alternatively, $C^{-1}.$ The obtained equations are valid in
the general case of large deformations and large deformation gradients.
The appropriate boundary conditions which complete the set of bulk
equations are also supplied%
\footnote{Some of the found equations are a possible regularization of those
proposed e.g. in Yeremeyev et al. \cite{EremeevFreidinSharipova2003,YeremeyevFreidinSharipova2007}
for phase transitions in solids and may give an insight into some
of the results presented in Eremeyev and Lebedev \cite{Eremeyev Lebedev2012}.%
}. The main computational tool that we use is the Levi-Civita tensor
calculus, also applied to embedded submanifolds. It has to be remarked
that the works of Piola, although correct and rigorous, are encumbered
by heavy component-wise notation hindered their comprehension. Piola's
works are truly modern in spirit, except in what concerns their difficulty
in treating tensorial quantities: the reader will appreciate the enormous
economy of thought which is gained by the use of Levi-Civita formalism.

Piola was aware of the difficulties which are to be confronted when
formulating new theories, as he claims (see \cite{Piola4}, page 1):

\textit{``It happens not so seldom that new achievements -by means
of which a branch of applied mathematics was augmented- do not appear
immediately, in the concept and in the exposition, free from lengthiness
and superfluence. The complication of analytical procedures can reach
such a level that it could seem impossible to proceed: indeed it is
in this moment instead that sometimes a more general point of view
can be discovered, many particularities are concentrated, and a compendious
theory is formed which is so well grounded that it can infuse vigor
for further progresses.''}

We conclude by citing a part of the Introduction of Piola \cite{Piola4},
page 5, which is suitable to include also the present one%
\footnote{The translation from the original italian text tries to reproduce
the English style of the famous works by Maxwell \cite{Maxwell},
which are nearly contemporary with Piola's ones.%
}, when decontextualizing the references to previous works and replacing
the word \emph{fluids} by \emph{capillary fluids}:

\textit{``While with the present memoir I will aim again to the goal
now devised}%
\footnote{\textit{\emph{Piola refers here to his intention of deducing all the
evolution equations of continuum mechanics from the principle of virtual
works.}}%
}\textit{ I will manage to reach also other ones. {[}Indeed{]} it is
rigorously proved in many places that the general equation of mechanics,
written with the notation of the calculus of variations, in the case
of a whatsoever discrete system of bodies regarded as points in which
different concentrated masses are subjected to external active forces
and to internal active and passive forces. However, to start from
this last equation {[}i.e. the equation for a discrete system of points{]}
and to obtain the formulas for the equilibrium and motion of bodies
with three dimensional extensions {[}i.e. deformable bodies{]}, it
indeed is a step very difficult for those who are willing to see things
clearly and who are not happy with an incomplete understanding. One
among my first efforts in this subject can be recognized in my Memoir
\textquotedbl{}On the principles of Analytical mechanics by Lagrange\textquotedbl{}.
Published in Milan already in the year 1825, where I presented in
this regard some correct ideas but with many specific technical details
either too complex or indeed superfluous. I came back to this point
in the memoir published in T. XXI of these Atti and I believed to
have obtained a remarkable improvement by introducing non-negligible
abbreviations and simplifications: but thereafter I perceived the
possibility of further improvements which I introduced in the present
one. Indeed great advantage can always be obtained when having the
care of clarifying appropriately the ideas concerning the nature of
different analytical quantities and the spirit of the methods: {[}to
establish{]} if also from this point of view something has been left
to be done, I will leave the judgment to intelligent readers. The
scholar will perceive that I propose myself also other aims with the
present work, having established here various formulas, which can
serve as a starting point for further investigations. I will not omit
to mention one of these aims and precisely that one which consists
in demonstrating anew (Capo V), by adopting the ideas better founded
which are provided by modern Physics about fluids, the fundamental
equations of their motions. Insofar as I treated at length in other
my works the problems of hydrodynamics (See the first two volumes
of Memoirs of I. R. Istituto Lombardo) it was objected that my deduction
could be defective, considering what was stated by Poisson about the
equations of ordinary Hydrodynamics. Now I believed to be able to
prove that the considerations of the French Geometer in this circumstance
were pushed too far ahead, and that notwithstanding his objections
the fundamental theory of the motions of fluids is well grounded as
established by D'Alembert and Euler, and exactly as it was reproduced
by Fourier himself with the addition of another equation deduced from
the theory of heat, {[}equation{]} to which, however, it is not necessary
to refer in the most obvious questions concerning the science of liquids.
For what concerns the motion of fluids, the present Memoir is intended
to support and complement the aforementioned ones.''}

In the Appendix B the reader will find various kinematic formulas,
which in our opinion will be useful in further developments of analytical
continuum mechanics. The reader should also explicitly note that already
Piola has stated that the heat equation does not need to be considered
when purely mechanical phenomena are involved.

\subsection{The meanings of the expressions: \emph{second gradient continua}
and\ \emph{capillary fluids}}

Following Germain \cite{GERMAIN1973} we will call second gradient
continua those media whose Lagrangian volumetric deformation energy
depends both on the first and second gradients of the placement field.
When using the expression \emph{capillary fluids} we will refer to
those continua whose Eulerian volumetric deformation energy density
depends both on their Eulerian mass density $\rho$ and its gradient
$\nabla\rho$. Of course the aforementioned dependencies must be independent
of the observer (this requirement was already demanded by Piola \cite{Piola2}).
We prefer to avoid naming the introduced class of fluids after Cahn
and Hilliard or Korteweg and de Vries, as done sometimes in the literature
(See e.g. Seppecher \cite{SEPPECHER1988,SEPPECHER1996,SEPPECHER1996b,SEPPECHER2000,SEPPECHER2001},
Casal and Gouin \cite{Casal-Gouin1985,CASALGOUIN1988}). This is done
in order to avoid ambiguities: Cahn and Hilliard, for instance, intended
the equations which were subsequently named after them to be valid
for the concentration of a solute in motion with respect to a stationary
solvent, and deduced them via \emph{molecular} arguments (modulo some
thermodynamically relevant terms, see Casal and Gouin \cite{Casal-Gouin1985}).
On the other hand the Korteweg-de Vries equations \cite{KortewegdeVries1895}
were originally deduced for a completely different class of phenomena:
waves on shallow water surfaces. Later it was discovered that they
can also be deduced from an \emph{atomistic} argument, since the so-called
Fermi-Pasta-Ulam \cite{FERMIPASTAULAM-1955} discrete system has Korteweg-De
Vries equations as its \emph{continuum limit}. Only in a later paper
(Korteweg \cite{KORTEWEG1901}) is a connection with capillarity phenomena
established.

The nomenclature \textit{capillary fluids} is suggestive of many of
the most fundamental phenomena which may be described by the model
discussed here: wettability, the formation of interfacial boundary
layers, the formation of liquid or gaseous films close to walls, the
formation and the motion of drops or bubbles inside another fluid
phase or the formation of pendant or sessile drops on a horizontal
plane and many others (see e.g. the papers by Seppecher \cite{SEPPECHER1988,SEPPECHER1996},
dell'Isola et al. \cite{dellisolagouinseppecher1995}, Gouin and Casal
\cite{Casal-Gouin1985}). Finally, we remark that second gradient
theories are strictly related to continuum theories with microstructure
(see e.g. Green, Rivlin \cite{GREENRIVLIN1964a,GREENRIVLIN1964b,GREENRIVLIN1964c,GREENRIVLIN1965},
Mindlin \cite{Mindlin1964,Mindlin1965,Mindlin1968}, Kroner \cite{Kroner1968}
and Toupin \cite{TOUPIN1962,TOUPIN1964}) as clarified in the note
by Bleustein \cite{Bleustein} and in the papers by Forest \cite{Forest-2009,Forest-2011}.

\section{An interlude: some apparent dichotomies}

\subsection{Analytical continuum mechanics and continuum thermodynamics}

It is natural here to refer to the original sources of analytical
continuum mechanics. Some of them are relatively close in time and,
very often, their spirit has been somehow misjudged. Sometimes they
were forgotten or considered by some authors as not being general
enough to found \emph{modern} mechanics. This is not our opinion.
However, instead of looking for new words to support this point of
view, we will continue to cite a champion of analytical mechanics:
the Italian mathematical-physicist Gabrio Piola. Despite his being
one of the founders of modern continuum mechanics, his contribution
to it has been seriously underestimated. To our knowledge the appropriate
expression \emph{analytical continuum mechanics} has not been considered
frequently up to now. In Maugin \cite{Maugin2000} the following statement
can be found

\textit{``}\emph{The road to the analytical continuum mechanics was
explored in particular by P.Germain \cite{GERMAIN1992}, but not in
a variational framework.''}

The concept underlying analytical continuum mechanics is opposed to
those underlying continuum thermodynamics. Actually continuum thermodynamics
is based on a postulation process which can be summarized as follows
(see e.g. Noll and Truesdell \cite{NollTruesdell}, Noll \cite{NollFoundation}): 
\begin{itemize}
\item find a set of kinematic fields of relevance in the formulation of
the considered continuum model which is sufficient to describe considered
phenomena; 
\item postulate a suitable number of balance laws having the structure of
conservation laws. Specify \emph{the physical} meaning of each conserved
quantity and introduce for each a \emph{flux}, a \emph{source} and
a \emph{volume density;} 
\item postulate a suitable number of constitutive equations required to
\emph{close} the formulated mathematical problems: that is to have
enough equations to determine the evolution of the kinematic fields,
once suitable initial and boundary data are assigned; 
\item as the possible choices of constitutive equations are too large, and
many of them are unphysical, choose a particular balance law, i.e.
the balance of entropy, and assume that its source is undeterminate
and always positive. The physically acceptable constitutive equations
are those for which all possible motions produce a positive entropy. 
\end{itemize}
Anyone who has carefully considered the efficacy of such an approach
may agree that: 
\begin{itemize}
\item when one wants to formulate new models it is difficult to use it as
a heuristic tool; 
\item it is somehow involved and often requires many \textit{ad-hoc} assumptions. 
\end{itemize}
A clear and elegant%
\footnote{and also relatively old: but older does not mean always worse! (see
\cite{Russo})%
}comparative analysis of the advantages obtained by using instead the
principle of virtual work (or the principle of least action) is found
in Hellinger \cite{Hellinger1914}. Actually a more elegant discussion
of this point can be found in Piola (Memoir \cite{Piola4} page 1)
where one can read the following words:

``{[}By means of the concepts of Analytical mechanics{]} \textit{a
compendious theory is formed which is so well-grounded that it can
infuse vigor for further progresses.} \textit{It should be desirable
that this could happen also for the last additions made by the modern
Geometers to Rational mechanics: and in my opinion I should say that
the true method suitable to succeed we have in our own hands: it has
to be seen if others will be willing to share my opinion.} I wrote
many times that it does not seem to me needed to create a new mechanics,
departing from the luminous method of Lagrange's Analytical mechanics,
if one wants to describe the internal phenomena occurring in the motion
of bodies: \textit{{[}indeed it is my opinion that{]} it is possible
to adapt those methods to all needs of modern Mathematical Physics
: {[}and that{]} this is, nay, the true route to follow because, being
well grounded in its principles, it leads to reliable consequences
and it promises ulterior and grandiose achievements. However I had
-and still nowadays I have- as opponents well respected authorities,
in front of whom I should concede the point, if the validity of a
scientific opinion had to be based on an argument concerning the scientific
value of its supporter. Nevertheless, as I cannot renounce to my persuasion,
I believed it was suitable to try another effort, gathering in this
memoir my thoughts about the subject and having care to expose them
with the accuracy needed to assure to them the due attention of Geometers.}
{[}...{]}\emph{ Even more than for its elegance and the grandiosity
of its analytical processes, the true reason for which I prefer to
all the other methods in mechanics those methods due to Lagrange is
that I see in them the expression of that wise inductive/deductive
philosophy brought to us by Newton, which starts from the facts to
rise up to the laws and then {[}starting from established laws{]}
goes down again to the explanation of other facts.}''

Indeed, analytical continuum mechanics has a much simpler postulation
process since one has to 
\begin{itemize}
\item postulate the form of a suitable action functional; 
\item postulate the form of a suitable dissipation Hamilton-Rayleigh functional,
and calculate its first variation with respect to velocity fields; 
\item assume that in conservative motions the action is stationary, and
to determine these motions by calculating the first variation of the
action and equating it to zero for every infinitesimal variation of
motion; 
\item equate, for non-conservative motions, the first variation of the action
functional (on the infinitesimal variations of motion) to the first
variation of the Hamilton-Rayleigh functional with respect to Lagrangian
velocities (estimated on the same infinitesimal variations of motion). 
\end{itemize}
The true difficulty in analytical continuum mechanics is that it strongly
relies on the methods and on the ideas of the calculus of variations.\textbf{
}Most likely it is for avoiding the mathematical abstraction required
by the calculus of variations that many opponents reject Lagrangian
mechanics. Again we give voice to Piola (\cite{Piola4} page 4):

\textit{``Somebody could here object that this }{[}i.e. the variational
foundations of Analytical mechanics{]} \textit{is a very old knowledge,
which does not deserve to be newly promulgated by me: however {[}it
seems that my efforts are needed{]} as my beautiful theories {[}after
being published{]} are then criticized, because Poisson has assured}\textbf{\textit{
}}\textit{us (Mémoires Institut de France T. VIII. pag. 326, 400;
Journal Ecole polyt. cah. XX. pag 2) that the Lagrangian method used
for writing the effects of the forces by means of constraint equations
(method which is proclaimed here as the only one really suited to
take into accounts facts instead of causes) is too abstract}\textbf{\textit{;
}}\textit{that it is necessary to develop a Science closer to the
reality of things; that such analysis\ {[}the Lagrangian one{]} extended
to the real bodies must be rejected as insufficient}\textbf{\textit{.}}\textit{
I respond that I also recognize the difficult question to be in these
considerations. If it is well founded or not the statement that the
Lagrangian methods are sufficient to the description of all Mechanical
Phenomena, and are so powerful that they are suitable for all further
possible researches, this is what will be decided later, and before
rebutting my point of view, it will be fair to allow me to expose
all arguments which I have gathered to defend my point of view. I
hope to clarify in the following Memoir that}\textbf{\textit{ }}\textit{the
only reason for which the Analytical Mechanics seemed to be insufficient
in the solution of some problems, is that Lagrange}\textbf{\textit{,}}\textit{
while writing the conditions for equilibrium and motion of a three
dimensional body, did not detail his model by assigning the equations
relative to every material point belonging to it. If he had done this}\textbf{\textit{,
}}\textit{and he could very well do it without departing from the
methods imparted in his book,}\textbf{\textit{ }}\textit{he would
have obtained easily the same equations at which the French Geometers
of our times arrived very painfully, {[}equations{]} which now are
the foundation of new theories. However those results which Lagrange
could not obtain, because death subtracted him prom science before
he could complete his great oeuvre, those results can be obtained
by others.''}

\subsection{Least (or stationary) action principle and the principle of virtual
work}

Let us consider a physical system denoted $\mathcal{S}$. The set
of the possible states this system is mathematically described by
a space of configurations $\mathcal{C}$. The time evolution of $\mathcal{S}$
is modeled by a suitably regular function of the time variable whose
values belong to $\mathcal{C}$. In the following this function will
be called motion function (or shortly: \emph{motion}). Therefore a
well-posed mathematical model for $\mathcal{S}$ can be specified
only by starting with the choice of a space of configurations and
a set of conditions which determine the motions.

\paragraph{Least action principle:}

\begin{center}
\textit{The motions in a time interval }$\left[t_{0},t_{1}\right]$\textit{\ can
be characterized as those motion functions which minimize } 
\par\end{center}

\begin{center}
\textit{(or which are stationary for) a suitably defined action functional
in a specified set of admissible motions. } 
\par\end{center}

Indeed it is very important, in order to have a well-posed minimization
(or stationarity) problem, to precisely specify the set of \emph{admissible}
motions among which these minimizers have to be sought. Following
Lagrange it is generally assumed that the set of admissible motions
is included in the set of isochronous motions between the instants
$t_{0}$ and $t_{1},$ i.e. motions which start from a given configuration
at instant $t_{0}$ and arrive to another given configuration at the
instant $t_{1}.$ When differential calculus is applicable to the
action functional, the first variation of this functional (in the
sense of Gâteaux derivative) can be estimated. This first variation
is a linear continuous functional defined on the set of isochronous
infinitesimal variations of motion. In this case, the stationarity
condition can be formulated as a differential equation. This equation
requires that the first variation to vanish for every infinitesimal
variation of motion.

Lagrange studied a particular class of action functionals and gave
a method for calculating their first variation under suitable regularity
conditions on the action functional and the admissible motions. The
resulting equations of motion are necessary and sufficient conditions
for the stationarity of a given action. This method allows for the
consideration of both finite- and infinite-dimensional configuration
spaces, hence the action principle can be formulated in both cases.
Lagrangian action functionals are given in terms of a suitable Lagrangian
function, whose integration in time (and also in space if the configuration
space is constituted by spatial fields) is required for calculating
the action relative to a given motion. The form of such a function
can be regarded as a conjectural choice, whose validity has to be
experimentally tested. One can say that a\textbf{ }\emph{constitutive}\textbf{
}choice is implicit in the choice of a Lagrange function.

However, given a configuration space $\mathcal{C}$, one can postulate,
instead of a least action principle, a principle of virtual work.
This principle states that the motion of the considered system is
characterized by assuming that for every (admissible) variation the
sum of three linear continuous functionals is vanishing. These functionals
are, respectively, the \emph{internal work}, the \emph{external work}
and the \emph{inertial work}. Their choice has a nature similar to
the one which leads to a Lagrangian function and is also conjectural
in nature. As previously, the validity of these \emph{constitutive}
equations has to be experimentally tested. It has to be remarked that
if a Lagrange action functional can be split into three parts, i.e.
into the sum of inertial, internal and external terms, the stationarity
of action implies the validity of a virtual work principle. However
it is clear that, in general, a linear continuous functional of infinitesimal
variations of motion is not the first variation of a functional necessarily.
In this sense the principle of virtual works is more general than
the principle of least action. The principle of virtual work includes
the principle of least action as modified by Hamilton and Rayleigh.

Therefore, and contrary to what is sometimes stated, both the principle
of least action and the principle of virtual work depend on fundamental
constitutive assumptions: those which lead to the choice of, respectively,
either the three work functionals or the Lagrangian function. The
principle of virtual works is, once the configuration space is fixed,
able to produce a wider class of motions. In particular it seems to
be able to describe a wider class of dissipative phenomena (see e.g.
Santilli \cite{Santilli}). However, it has to be remarked that

i) there are dissipative systems which are governed by a least action
principle (see e.g. Moiseiwitsch \cite{Moisewitsch} or Vujanovic
and Jones \cite{VujanovicJones1989});

ii) it is conceivable, by a suitable embedding into a larger space
of configuration, to find Lagrangian forms for systems which are initially
not Lagrangian (see again Santilli \cite{Santilli} or Carcaterra
and Akay \cite{CarcaterraAkay2007,CarcaterraAkay2011}).

The physical insight gained using the principle of least action (or
the principle of virtual work) cannot be over estimated. For a deeper
discussion of this point we limit ourselves to cite here, among the
vast literature, the textbooks Landau and Lifshitz \cite{Landau1977},
Lanczos \cite{Lanczos}, Soper \cite{Soper}, Bedford \cite{Bedford1985},
Kupershmidt \cite{Kupershmidt1992}, Kravchuk and Neittaanmaki \cite{KRAVCHUKNEITTAANMAKI2007},
Lemons D.S. \cite{Lemons1997} and the methodological essay by Edwards
\cite{Edwards}. Some results of interest in continuum mechanics and
structural engineering are gathered in Leipholz \cite{Leipholz},
and Lippmann \cite{Lippmann1972}, while in Luongo and Romeo \cite{LuongoRomeo2006},
Luongo et al. \cite{LuongoDi Egidio2005,LuongoZulliPiccardo2009},
are presented some interesting results in the nonlinear dynamics of
some structural members.

\subsection{Discrete and continuous models}

In many works (see e.g. Truesdell \cite{Truesdell1968}) it is stated
that the principle of least action is suitable to derive the evolution
equations for finite dimensional systems only. Moreover, in some \emph{époques}
and some cultural environments, the atomistic vision prevailed in
physics to the extent that continuum models were considered inappropriate
simply for philosophical reasons. Indeed already Poisson bitterly
criticized the first works of Piola (see e.g. the introduction of
\cite{Piola4}) in which the foundations of modern continuum mechanics
are laid based on the principle of virtual work. Actually in Poisson's
opinion the true physical reality was \emph{atomistic} and the most
fundamental concept in mechanics was the concept of force, whose balance
was bound to lead to the evolution equations of every mechanical system.
As a consequence and in order to respond to the objections of Poisson,
even if Piola was aware that a variational deduction of the evolution
equations for continuous systems was possible, in the first half of
the XIX century he decided to regard the continuum theory as the limit
of a discrete system. It is interesting to remark that, only a few
years later, a similar controversy arose between Mach and Boltzmann,
based on Mach's rejection of the atomistic point of view in thermodynamics.
We prefer to leave Piola (\cite{Piola4} page 2) to explain his (and
our) point of view:

\textit{``In my opinion it is not safe enough to found the primordial
formulas {[}of a theory{]} upon hypotheses which, even being very
well-thought, do not receive support if not for a far correspondence
with some observed phenomena, correspondence obtained by particularizing
general statements, {[}in my opinion{]} this should be as coming back
in a certain sense to the philosophy of Descartes and Gassendi: indeed
the }\textit{\emph{magisterium}}\textit{ }\textit{\emph{of nature}}\textit{
{[}the experimental evidence{]} at the very small scale in which we
try to conceive the effect of molecular actions will perhaps actually
be very different from what we can mentally realize by means of the
images impressed in our senses when experiencing their effects on
a larger scale. Even let us assume that this difference be very small:
a deviation quite insensitive in the fundamental constituents {[}of
matter{]} -which one needs to consider as multiplied by millions and
by billions before one can reach sensible dimensions- can be the ultimate
source of notable errors. On the contrary, by using Lagrangian methods,
one does not consider in the calculations the actions of internal
forces but {[}only{]} their effects, which are well-known and are
not at all influenced by the incertitude about the effects of prime
causes,\ {[}so that{]} no doubt can arise regarding the exactitude
of the results. It is true that our imagination may be less satisfied,
as {[}with Lagrangian methods{]} we do not allow to it to trace the
very fundamental origins of the internal motions in bodies: does it
really matter? A very large compensation for this deprivation can
be found in the certitude of deductions. I could here repeat, if they
were not very well-known, the wise documents with which Newton summoned
to the science of facts those philosophers who before him had left
a too free leap to their imagination. It has to be remarked that I
do not intend for this reason to proscribe the dictation of modern
Physics about the internal constitution of bodies and the molecular
interactions; I think, nay, to render to them the greater of services.
When the equations of equilibrium and motion will be established firmly
upon indisputable principles, because one has calculated certain effects
rather than hypothetical expression of forces, I believe to be licit
to try to reconstruct anew these equations by means of {[}suitable{]}
assumptions about such molecular interactions: and if we manage in
this way to get results which are identical to those we already know
to be true, I believe that these hypotheses will acquire such a high
degree of likeliness which one could never hope to get with other
methods.}\textbf{\ }\textit{Then the molecular Physics will be encouraged
to continue with its deductions, under the condition that, being aware
of the aberrations of some bald ancient thinkers, it will always mind
to look carefully in the experimental observation those hints {[}coming
by the application of Lagrangian macroscopic methods{]} which are
explicit warnings left there to indicate every eventual deviation.''}

Regarding the concept of characteristic scale lengths relevant in
physical phenomena, Piola had crystal clear ideas, expressed by him
with such an elegance that even nowadays his words can be used (Piola
\cite{Piola4} page 13):

\textit{``Scholium. The admissibility of the principle }{[}i.e. the
principle which assumes the existence of a characteristic length $\sigma$
determining the \emph{average} distance among the molecules microscopically
constituting the considered continuum{]} \textit{refers to the true
condition of the human being, placed, as said by Pascal in his Thoughts
(Part I. Art.IV) at immense distances both from infinity and the zero:
distances in which one can imagine many orders of magnitude, of which
one {[}order of magnitude{]} can be regarded as the whole when compared
with the one which is preceding it, and nearly nothing when compared
with {[}the order of magnitude{]} which follows it. Therefore it results
that the same quantities which are asserted to be negligible for us
without being afraid of being wrong, could be great and not at all
negligible quantities for beings which could be, for instance, capable
to perceive the proportions which are relevant for the structure of
}\textit{\emph{micro-organisms}}\textit{ . For those beings those
bodies which appear to us to be continuous could appear as bunches
of sacks: water, which for us is a true liquid, could appear as for
us {[}appears{]} millet or a flowing bunch of lead pellets. But also
for these beings there would exist true fluids, relative to which
for them the same consequences which we deduce relatively to water
should be considered as true. There are therefore quantities which
are null absolutely for all orders of beings, as the analytical elements
used in the Integral Calculus, and there are quantities which are
null only for beings of a certain order, and these quantities would
not be null for other beings, as some elements which are considered
in mechanics. As I was educated by Brunacci to the school of Lagrange,
I always opposed to the metaphysical infinitesimal, as I believe that
for the analysis and the geometry (if one wants to achieve clear ideas)
it has to be replaced by the }\textit{\emph{indeterminately}}\textit{
}\textit{\emph{small}}\textit{ when it is needed:\ however I accept
the physical infinitesimal, of which the idea is very clear. It is
not an absolute zero, it is nay a magnitude which for other beings
could be appreciable, but it is a zero relative to our senses, for
which everything which is below them is exactly as if it were not
existing.}

The reader should remark that the original formulations which lead
to the Cahn-Hilliard equations \cite{Cahn-Hilliard1958,Cahn-Hilliard1959}
and to capillary fluid equations (see e.g. van Kampen \cite{VANKAMPEN1964},
Evans \cite{EVANS1979}, De Gennes \cite{DEGENNES1981}) were based
on atomistic arguments. However these arguments may lead sometimes
to equations (see for more details Casal and Gouin \cite{Casal-Gouin1985})
which are thermodynamically inconsistent. This circumstance was already
clear to Piola, who suggested the use of macroscopic theories (based
on the principle of virtual work) to \emph{derive} and \emph{confirm}
the correct deductions from atomistic arguments. This good scientific
practice is nowadays generally accepted. Many efforts have been dedicated
to deduce from an \emph{atomistic} scale discrete model the \emph{macroscopic}
form of the deformation energies which depend on first or higher gradients
of deformation starting from the works of Piola \cite{Piola4}. The
reader is referred to Esposito and Pulvirenti \cite{EspositoPulvirenti2004}
for an extensive review about the results available for fluids. It
is suggestive to conjecture that the macro-models for fluid flows
discussed e.g. in \cite{Bassanini et al 1996,Casciola et al. 2005,JacobCasciolaetal2008},
which involve some micro-macro identification procedure and more than
one length scale, may be framed in the general scheme which is put
forward here. In solid mechanics also, multiscale models have attracting
the interest of many authors: we may refer, for instance, to Sunyk
and Steinmann \cite{SunykSteinmann2003}, Alibert et al. \cite{AlibertSeppecherdellisola},
Steinmann et al. \cite{SteinmannElizondoSunyk2007}, Rinaldi et al.
\cite{Rinaldietal2008}, Misra and Chang \cite{MISRA1}, Yang and
Misra \cite{YangMisra2010}\cite{YANGMISRA2012}, Yang et al. \cite{YangMisra2011},
Misra and Singh \cite{MISRA3}, Misra and Ching \cite{MISRA4} for
some other interesting results concerning granular solids. In the
same context the results presented in Boutin and Hans \cite{Boutin3},
Auriault et al. \cite{Auriault}, Chesnais et al. \cite{Chesnais,Chesnaisetal},
Soubestre \cite{Soubestre} and Boutin \cite{Boutin2,Boutin1} have
also to be cited. In these papers the authors, although starting in
their procedure from balance laws valid at a microscopic level, proceed
in a spirit very similar to the one found in the pioneering works
by Piola.

\part{Deduction of evolution equations for continuous systems using the
least action principle}

In this part, starting from the least action principle, we present
the formal deduction of the evolution equations which govern the motion
of i) first gradient continua, in particular Euler fluids, and of
ii) second gradient continua, in particular capillary fluids. Although
the content of the following subsection is well-known (even if more
or less consciously ignored in some literature) it was written pursuing
a twofold aim: i) to establish the notation and calculation tools
to be used in the subsequent sections; ii) to rephrase there, in a
modern notation, the results of Piola \cite{Piola1,Piola5}. It has
to be remarked that in the literature the least action principle in
continuum mechanics is presented in a very clear way in Berdichevsky
\cite{Berdi}. It is evident that the Soviet school (see e.g. Sedov
\cite{Sedov,Sedov Vienna}, which developed, improved and elaborated
it in several aspects), was aware of the content of Piola's contribution
to continuum mechanics%
\footnote{It cannot be excluded logically that Piola could have sources which
we could not find. However his works fix a date from which certain
concepts start to appear in published-printed form.%
}, even if it is not so clear how the information managed to reach
Soviet scientists. To establish the ways in which such connections
are established is a scientific problem by itself, whose importance
has been underestimated up to now.

\section{First gradient continua}

In this section we reproduce, by introducing more recent notations
and by extensively using Levi-Civita's absolute tensor calculus, the
arguments used by Piola for founding the classical continuum mechanics.
The reader will observe by simple comparison (see Piola \cite{Piola1,Piola2,Piola3,Piola4,Piola5})
that the use of tensor calculus makes the presentation dramatically
shorter. Moreover, as we will see in a subsequent subsection, by means
of its use the calculations needed to deal with second gradient fluids
become feasible. Another difference with Piola's presentation consists
in our use of the least action principle instead of the principle
of virtual work (see e.g. dell'Isola and Placidi \cite{dellisolaplacidi2011}).
However we keep the distinction among inertial, internal and external
actions. Notations used in the following are detailed in the Appendices.

\subsection{Action functional}

Let us introduce the following action functional: 
\[
\mathcal{A}=\int_{t_{o}}^{t_{1}}\int_{\mathcal{B}}\left(\frac{1}{2}\rho_{0}v^{2}-W(\chi,F,X)\right)dVdt+\int_{t_{o}}^{t_{1}}\int_{\partial\mathcal{B}}\left(-W_{\mathrm{S}}(\chi,X)\right)dAdt
\]
where : 
\begin{itemize}
\item the field $\chi$ denotes the placement field between the referential
(or Lagrangian) $\mathcal{B}$ and the spatial (or Eulerian) $\chi\left(\mathcal{B}\right)\subset\mathcal{E}$
configurations 
\[
\chi:\mathcal{B}\rightarrow\mathcal{E}
\]

\item the field $\rho_{0}(X)$ refers to the Lagrangian time-independent
mass density, so that the Eulerian mass density is given by 
\[
\rho=\det F^{-1}\left(\rho_{0}\right)^{\overrightarrow{(\mathcal{E})}}
\]
where the used notation is carefully defined in Appendix A; 
\item the placement gradient $F=\nabla_{X}\chi$ is a Lagrangian tensor
field, i.e. a tensor field defined on $\mathcal{B}$; 
\item the velocity field $v=\frac{\partial\chi}{\partial t}$ , associated
to the placement field $\chi,$ is a Lagrangian field of Eulerian
vectors; 
\item the potential $W(\chi,F,X)$ is relative to the volumetric density
of action inside the volume $\mathcal{B}$; 
\item the potential $W_{S}(\chi,X)$ pertains to the actions externally
applied at the boundary $\partial\mathcal{B}$. 
\end{itemize}
The results valid for infinite dimensional Lagrangian models (see
e.g. dell'Isola and Placidi \cite{dellisolaplacidi2011} and references
therein) applied to the introduced action, lead to the following Euler-Lagrange
equations (which hold at every internal point of $\mathcal{B}$):
\[
-\frac{\partial}{\partial t}\left(\rho_{0}v_{i}\right)+\frac{\partial}{\partial X^{A}}\left(\frac{\partial W}{\partial F_{A}^{i}}\right)-\frac{\partial W}{\partial\chi^{i}}=0
\]
and, if the boundary $\partial\mathcal{B}$ is suitably smooth, the
following boundary conditions%
\footnote{To avoid any misunderstanding, in the expression $W_{\mathrm{S}}$
the subscript $S$ refers to ``Surface'' and is not an index.%
} 
\[
-\frac{\partial W}{\partial F_{A}^{i}}N_{A}-\frac{\partial W_{\mathrm{S}}}{\partial\chi^{i}}=0.
\]
which hold at every point $P$ belonging to the (Lagrangian) surface
$\partial\mathcal{B}$ whose normal field is denoted $N(P)$ or, in
components, $N_{M}(P)$. In the former expressions and throughout
the paper, Lagragian indices are written in upper case while Eulerian
indices are written in lower case. Furthermore the classical Einstein
convention is applied and the summed indices are taken from the beginning
of the alphabet.

\subsection{Objective deformation energy}

We now assume that the energy $W$ can be split into two parts, the
first one representing the deformation energy, the second one an external
(conservative) action of a bulk load 
\[
W(\chi,F,X)=W^{\mathrm{def}}(C,X)+U^{\mathrm{ext}}(\chi,X)
\]
where $C:=F^{T}F$ is the right Cauchy-Green tensor which, in components,
has the following expressions: 
\[
C_{MN}=g_{NA}F_{a}^{A}F_{M}^{a}=F_{Na}F_{M}^{a}=g_{ab}F_{M}^{b}F_{N}^{a},
\]
where $g_{MN}$ and $g_{ij}$ denotes, respectively, the (distinct)
metric tensors over $\mathcal{B}$ and $\mathcal{E}$. The Euler-Lagrange
stationarity conditions are the so-called \emph{balance of linear
momentum,} or \emph{balance of forces,} represented by the equations
\begin{equation}
-\frac{\partial}{\partial t}\left(\rho_{0}v_{i}\right)+\frac{\partial}{\partial X^{C}}\left(\frac{\partial W^{\mathrm{def}}}{\partial C_{AB}}\frac{\partial C_{AB}}{\partial F_{C}^{i}}\right)-\frac{\partial U^{\mathrm{ext}}}{\partial\chi^{i}}=0.\label{BalanceForce Lagrangian}
\end{equation}
Observe that the equality concerns Eulerian vectors, but the fields
are expressed in terms of the Lagrangian variables; therefore the
differential operators are Lagrangian. Let us now observe that as:
\[
\frac{\partial C_{MN}}{\partial F_{P}^{i}}=g_{ab}\frac{\partial}{\partial F_{P}^{i}}\left(F_{M}^{b}F_{N}^{a}\right)=g_{ab}\left(\frac{\partial F_{M}^{b}}{\partial F_{P}^{i}}F_{N}^{a}+F_{M}^{b}\frac{\partial F_{N}^{a}}{\partial F_{P}^{i}}\right)=\left(\delta_{M}^{P}F_{iN}+F_{iM}\delta_{N}^{P}\right)
\]
we get 
\[
\frac{\partial W^{\mathrm{def}}}{\partial C_{AB}}\frac{\partial C_{AB}}{\partial F_{P}^{i}}=2\frac{\partial W^{\mathrm{def}}}{\partial C_{PA}}F_{iA}
\]
and the balance \eqref{BalanceForce Lagrangian} becomes 
\begin{equation}
-\rho_{0}\frac{\partial v_{i}}{\partial t}+\frac{\partial}{\partial X^{A}}\left(2F_{iB}\frac{\partial W^{\mathrm{def}}}{\partial C_{AB}}\right)-\frac{\partial U^{\mathrm{ext}}}{\partial\chi^{i}}=0.\label{Balance Lagrangian}
\end{equation}
The tensor 
\[
P_{i}^{M}:=2F_{iA}\frac{\partial W^{\mathrm{def}}}{\partial C_{AB}}g^{BM}
\]
is the Piola stress tensor. It appears also in the boundary conditions
which are deduced from 
\begin{equation}
\frac{\partial W^{\mathrm{def}}}{\partial F_{A}^{i}}N_{A}=-\frac{\partial W_{S}}{\partial\chi^{i}}\label{BoundaryConditionLagrangian}
\end{equation}
In Piola \cite{Piola4} the requirement of objectivity (i.e. the invariance
under changes of observer) of Piola stress is clearly stated and analytically
formulated. However, due to the lack of conceptual tools supplied
by tensor calculus, in his results he did not achieve the clarity
allowed by the tensorial formalism.

\subsection{The Eulerian form of force balance}

Using the Piola transformation (see Appendices), the equations \eqref{Balance Lagrangian},
which represent the equations of motion become 
\[
-\left(\rho_{0}\left.\frac{\partial v_{i}}{\partial t}\right\vert _{X}\right)^{\overrightarrow{(\mathcal{E})}}+J^{\overrightarrow{(\mathcal{E})}}\frac{\partial}{\partial x^{a}}\left(2J^{-1}\left(F_{iA}\frac{\partial W^{\mathrm{def}}}{\partial C_{AB}}F_{B}^{a}\right)^{\overrightarrow{(\mathcal{E})}}\right)-\left(\frac{\partial U^{\mathrm{ext}}}{\partial\chi^{i}}\right)^{\overrightarrow{(\mathcal{E})}}=0.
\]
We remark here that $J^{-1}=\det\left(F^{-1}\right)$, consequently
$J^{-1}$ has to be considered as an Eulerian quantity. Multiplying
this expression by $J^{-1}$ one gets 
\begin{equation}
-J^{-1}\left(\left.\rho_{0}\frac{\partial v_{i}}{\partial t}\right\vert _{X}\right)^{\overrightarrow{(\mathcal{E})}}+\frac{\partial}{\partial x^{a}}\left(2J^{-1}\left(F_{iA}\frac{\partial W^{\mathrm{def}}}{\partial C_{AB}}F_{B}^{a}\right)^{\overrightarrow{(\mathcal{E})}}\right)-J^{-1}\left(\frac{\partial U^{\mathrm{ext}}}{\partial\chi^{i}}\right)^{\overrightarrow{(\mathcal{E})}}=0.\label{Eulerian Balance of Force}
\end{equation}
These are recognized as the celebrated balance equations of linear
momentum of classical continuum mechanics, once one introduces: 
\begin{enumerate}
\item the Cauchy stress tensor (which is self-adjoint) 
\begin{equation}
T_{i}^{j}:=2J^{-1}\left(F_{iA}\frac{\partial W^{\mathrm{def}}}{\partial C_{AB}}F_{B}^{j}\right)^{\overrightarrow{(\mathcal{E})}}\label{eq:cauchy}
\end{equation}

\item the material Eulerian time derivative of a Lagragian field $\Phi$
as 
\[
\left(\left.\frac{\partial\Phi}{\partial t}\right\vert _{X}\right)^{\overrightarrow{(\mathcal{E})}}
\]

\item the field 
\[
b^{\mathrm{ext}}:=-J^{-1}\left(\frac{\partial U^{\mathrm{ext}}}{\partial\chi^{i}}\right)^{\overrightarrow{(\mathcal{E})}}
\]
which can be called the Eulerian volume force density for considered
bulk loads. 
\end{enumerate}
Finally, to transport the boundary conditions \eqref{BoundaryConditionLagrangian}
into the Eulerian configuration, we introduce the following notations,
assumptions and results: 
\begin{enumerate}
\item The body boundary $\partial\mathcal{B}$, whose unit normal field
is denoted $N$, is mapped by the placement $\chi$ onto the Eulerian
surface $\chi(\partial\mathcal{B})$ whose unit normal field is denoted
$n$; 
\item Particularizing the relations \eqref{TRANSFORMATIONOFNORMALS} and
\eqref{AREASTRANFORMATION} provided in the appendix, we obtain that
\begin{equation}
N_{M}^{\overrightarrow{(\mathcal{E})}}=\frac{\left(J^{-1}\left(F^{T}\right)_{M}^{a}\right)n_{a}}{\left\Vert \left(J^{-1}\left(F^{T}\right)_{A}^{b}\right)n_{b}\right\Vert }\label{NORMALSEULERLAGRANGE}
\end{equation}
and that 
\begin{equation}
\frac{dA_{\mathcal{E}}}{dA_{\mathcal{B}}}=\left(\left\Vert \left(J^{-1}\left(F^{T}\right)_{A}^{a}\right)n_{a}\right\Vert ^{-1}\right)^{\overrightarrow{(\mathcal{B})}}=\left\Vert \left(J\left(F^{-T}\right)_{a}^{A}\right)N_{A}\right\Vert \label{AREAS}
\end{equation}

\item The Lagrangian conditions \eqref{BoundaryConditionLagrangian} imply
\[
2\frac{\partial W^{\mathrm{def}}}{\partial C_{AB}}F_{iA}N_{B}=-\frac{\partial W_{S}}{\partial\chi^{i}}
\]
which, by using \eqref{NORMALSEULERLAGRANGE}, become 
\[
2F_{iA}\frac{\partial W^{\mathrm{def}}}{\partial C_{AB}}F_{B}^{a}\left(J^{-1}n_{a}\right)^{\overrightarrow{\left(\mathcal{B}\right)}}=-\frac{\partial W_{S}}{\partial\chi^{i}}\left\Vert \left(J\left(F^{-T}\right)_{a}^{A}\right)N_{A}\right\Vert 
\]

\end{enumerate}
These last equations, by using \eqref{eq:cauchy} and \eqref{AREAS},
allow us to obtain the well-known Eulerian boundary conditions 
\begin{equation}
T_{i}^{a}n_{a}=\left(-\frac{dA_{\mathcal{E}}}{dA_{\mathcal{B}}}\frac{\partial W_{S}}{\partial\chi^{i}}\right)^{\overrightarrow{(\mathcal{E})}}\label{CauchyTractionCondition}
\end{equation}

\subsection{Euler fluids}

We now continue to parallel Piola (\cite{Piola4} Capo V pages 111-146).
However our treatment differs since we characterize the material symmetry
of Euler fluids by assuming the equation \eqref{EnergyEulerFluids}
below, while Piola imposes it on the Eulerian transformation of Piola
stress. Let us assume that 
\begin{equation}
W^{\mathrm{def}}(C)=\Psi(\rho^{\overrightarrow{\left(\mathcal{B}\right)}}(C))=W^{\mathrm{eul}}(F)\label{EnergyEulerFluids}
\end{equation}
and recall the following relations: 
\[
\rho^{\overrightarrow{\left(\mathcal{B}\right)}}=\rho_{0}\left(\det F\right)^{-1};\qquad(\det F)^{2}=\det(F^{T}F)=\det(C);\qquad\rho^{\overrightarrow{\left(\mathcal{B}\right)}}=\rho_{0}\left(\det C\right)^{-\frac{1}{2}}
\]
To particularize \eqref{Eulerian Balance of Force} we need to determine
the special form assumed by Cauchy's tensor for Euler fluids. This
is done by using: 
\begin{enumerate}
\item The equality \eqref{DERIVATIVEMASSGREEN} given in the appendices
\[
\frac{\partial\rho^{\overrightarrow{\left(\mathcal{B}\right)}}}{\partial C_{MN}}=-\frac{\rho^{\overrightarrow{\left(\mathcal{B}\right)}}}{2}\left(F^{-1}\right)^{Ma}\left(F^{-1}\right)_{a}^{N}
\]

\item The equality 
\begin{align}
T_{i}^{j} & =2J^{-1}\left(F_{iA}\frac{\partial\Psi}{\partial\rho^{\overrightarrow{\left(\mathcal{B}\right)}}}\frac{\partial\rho^{\overrightarrow{\left(\mathcal{B}\right)}}}{\partial C_{AB}}F_{B}^{j}\right)^{\overrightarrow{\left(\mathcal{E}\right)}}=-J^{-1}\rho\frac{\partial\Psi}{\partial\rho}\delta_{i}^{a}\delta_{a}^{j}=-\rho^{2}\frac{\partial\left(\Psi/\rho_{0}^{\overrightarrow{\left(\mathcal{E}\right)}}\right)}{\partial\rho}\delta_{i}^{j}\label{CauchyFluids}
\end{align}

\item The definition of the constitutive equation giving the pressure as
a function of density 
\begin{equation}
p(\rho):=\rho^{2}\frac{\partial\left(\Psi/\rho_{0}^{\overrightarrow{\left(\mathcal{E}\right)}}\right)}{\partial\rho}\label{Pressure}
\end{equation}

\end{enumerate}
In conclusion, by using \eqref{Pressure} and \eqref{CauchyFluids},
the Eulerian force balance equations assume the form: 
\[
-\rho_{0}^{\overrightarrow{(\mathcal{E})}}J^{-1}\left(\left.\frac{\partial v_{i}}{\partial t}\right\vert _{X}\right)^{\overrightarrow{(\mathcal{E})}}-\frac{\partial p(\rho)}{\partial x^{i}}-J^{-1}\left(\frac{\partial U}{\partial\chi^{i}}\right)^{\overrightarrow{(\mathcal{E})}}=0.
\]
By considering the external potential energy per unit mass, the last
equation reads 
\[
-\rho\left(\left.\frac{\partial v_{i}}{\partial t}\right\vert _{X}\right)^{\overrightarrow{(\mathcal{E})}}-\frac{\partial p(\rho)}{\partial x^{i}}-\rho\left(\frac{\partial\left(U/\rho_{0}\right)}{\partial\chi^{i}}\right)^{\overrightarrow{(\mathcal{E})}}=0.
\]
Finally, using the formula for calculating the material derivative
of velocity we obtain 
\[
-\rho\left(\frac{\partial v_{i}^{\overrightarrow{(\mathcal{E})}}}{\partial t}+\frac{\partial v_{i}^{\overrightarrow{(\mathcal{E})}}}{\partial x^{a}}\left(v^{a}\right)^{\overrightarrow{(\mathcal{E})}}\right)-\frac{\partial p(\rho)}{\partial x^{i}}-\rho\left(\frac{\partial\left(U/\rho_{0}\right)}{\partial\chi^{i}}\right)^{\overrightarrow{(\mathcal{E})}}=0.
\]
The expression \eqref{CauchyFluids} for the Cauchy stress, which
is valid for Euler fluids, together with the boundary condition \eqref{CauchyTractionCondition}
implies that:\medskip{}

\begin{center}
\textbf{\textsl{Not all externally applied actions can be sustained
by Euler fluids. Indeed Euler fluids cannot sustain arbitrary surface
tractions (as pressure is always positive) nor surface shear forces.
}} 
\par\end{center}

\medskip{}

\noindent This statement, which can be already found in Piola \cite{Piola4}
(see equation (37) page 136 and the subsequent discussion), implies
that:

\medskip{}

\noindent \textit{``The assumptions about the internal deformation
energy determine the capability of the considered body to sustain
externally applied actions. Therefore: the expression of the internal
deformation energy characterizes the class of admissible external
actions of a continuous body.''}

\medskip{}

\noindent We will return on this point in the next sections.

\section{Second gradient continua}

In this section we generalize the expression for deformation energy
used up to now to take the second gradient of the displacement field
into account. It has to be remarked that in Piola (\cite{Piola4}
page 152) a first (and persuasive!) argument supporting the possible
importance of dependence of the internal work functional on higher
gradients of displacement field is put forward. This point deserves
a deeper discussion and is postponed to future investigations. To
our knowledge Piola is the first author who analyzed such a dependence.
Therefore we propose to name after him the obtained generalized continuum
theories. It is assumed that second gradient materials have a deformation
energy which depends both on the Cauchy-Green tensor and on its first
gradient. The more general Lagrangian density function to be considered
has the following form 
\begin{equation}
\mathcal{L}=\frac{1}{2}\rho_{0}v^{2}-(W^{I}(\chi,F,X)+W^{II}(\chi,F,\nabla F,X)).\label{GeneralizedLagrangian}
\end{equation}

\subsection{Piola-type second gradient deformation energy}

The expression \eqref{GeneralizedLagrangian} will be assumed in the
sequel. The term $W^{I}(\chi,F,X)$ coincides with the first order
term previously considered, while $W^{II}(\chi,F,\nabla F,X)$ stands
for an additive term in which the first order derivative of the gradient
$F$ appears. As a consequence, we need to compute the first variation
of the following functional 
\[
\mathcal{A}^{II}=\int_{\mathcal{B}}-W^{II}(\chi,F,\nabla F,X)dV.
\]
Paralleling the style of presentation used by Piola, while developing
the calculations we comment on the results as soon as they are obtained.
Because of the assumed structure of the added deformation energy,
we have 
\begin{align*}
\delta\mathcal{A}^{II} & =\delta_{\chi}\mathcal{\mathcal{A}}^{II}+\delta_{F}\mathcal{\mathcal{A}}^{II}+\delta_{\nabla F}\mathcal{\mathcal{A}}^{II}\\
 & =\int_{\mathcal{B}}-\left(\frac{\partial W^{II}(\chi,F,\nabla F,X)}{\partial\chi}\delta\chi+\frac{\partial W^{II}(\chi,F,\nabla F,X)}{\partial F}\delta F+\frac{\partial W^{II}(\chi,F,\nabla F,X)}{\partial\nabla F}\delta\nabla F\right)dV
\end{align*}
It can be observed that the first two terms can be treated exactly
in the manner of the first gradient action. The following expression
in the bulk equation will be obtained 
\begin{equation}
DIV_{X}\left(\frac{\partial W^{II}}{\partial F}\right)-\frac{\partial W^{II}}{\partial\chi}\label{DAF Bulk}
\end{equation}
together with the following term in the boundary conditions: 
\begin{equation}
-\frac{\partial W^{II}}{\partial F}\cdot N.\label{DAF Surface}
\end{equation}
On the contrary, new difficulties appear when calculating the first
variation $\delta_{\nabla F}\mathcal{\mathcal{A}}.$ However, the
techniques developed by Mindlin, Green, Rivlin, Toupin and Germain
(see also dell'Isola et al. \cite{dellisolaseppechermadeo2012}) allow
us to treat this term efficiently and elegantly. Starting from (the
comma indicates partial differentiation)%
\footnote{In this paper we introduce in both referential and spatial configurations
a chart with vanishing Christoffel symbol so that covariant derivatives
coincide with derivatives.%
} 
\[
\delta_{\nabla F}\mathcal{\mathcal{A}}^{II}=\int_{\mathcal{B}}-\left(\frac{\partial W^{II}}{\partial F_{A,B}^{a}}\delta F_{A,B}^{a}\right)dV
\]
we perform a first integration by parts. Indeed remarking that 
\[
\frac{\partial}{\partial X^{B}}\left(\frac{\partial W^{II}}{\partial F_{A,B}^{a}}\delta F_{A}^{a}\right)=\frac{\partial}{\partial X^{B}}\left(\frac{\partial W^{II}}{\partial F_{A,B}^{a}}\right)\delta F_{A}^{a}+\frac{\partial W^{II}}{\partial F_{A,B}^{a}}\delta F_{A,B}^{a}
\]
and applying the divergence theorem (recall that we denote by $N_{M}$
the components of the unit normal to the surface $\partial\mathcal{B}$),
we obtain 
\begin{align}
\delta_{\nabla F}\mathcal{\mathcal{A}}^{II} & =\int_{\mathcal{B}}-\left(\frac{\partial W^{II}}{\partial F_{A,B}^{a}}\delta F_{A,B}^{a}\right)dV=-\int_{\mathcal{B}}\frac{\partial}{\partial X^{B}}\left(\frac{\partial W^{II}}{\partial F_{A,B}^{a}}\delta F_{A}^{a}\right)dV+\int_{\mathcal{B}}\left(\frac{\partial}{\partial X^{B}}\left(\frac{\partial W^{II}}{\partial F_{A,B}^{a}}\right)\delta F_{A}^{a}\right)dV\nonumber \\
 & =-\int_{\partial\mathcal{B}}\left(\frac{\partial W^{II}}{\partial F_{A,B}^{a}}N_{B}\right)\delta F_{A}^{a}dA+\int_{\mathcal{B}}\left(\frac{\partial}{\partial X^{B}}\left(\frac{\partial W^{II}}{\partial F_{A,B}^{a}}\right)\delta F_{A}^{a}\right)dV.\label{DADF FINAL}
\end{align}
Let us observe that the second term of the previous expression has
exactly the same form as the first variation of the first gradient
action. Therefore this term becomes 
\begin{align*}
\int_{\mathcal{B}}\left(\frac{\partial}{\partial X^{B}}\left(\frac{\partial W^{II}}{\partial F_{A,B}^{a}}\right)\delta F_{A}^{a}\right)dV & =\int_{\mathcal{B}}\frac{\partial}{\partial X^{A}}\left(\frac{\partial}{\partial X^{B}}\left(\frac{\partial W^{II}}{\partial F_{A,B}^{a}}\right)\delta\chi^{a}\right)dV-\int_{\mathcal{B}}\left(\frac{\partial^{2}}{\partial X^{A}\partial X^{B}}\left(\frac{\partial W^{II}}{\partial F_{A,B}^{a}}\right)\delta\chi^{a}\right)dV\\
 & =\int_{\partial\mathcal{B}}\left(N_{A}\frac{\partial}{\partial X^{B}}\left(\frac{\partial W^{II}}{\partial F_{A,B}^{a}}\right)\right)\delta\chi^{a}dA-\int_{\mathcal{B}}\left(\frac{\partial^{2}}{\partial X^{A}\partial X^{B}}\left(\frac{\partial W^{II}}{\partial F_{A,B}^{a}}\right)\delta\chi^{a}\right)dV
\end{align*}
which implies that to \eqref{DAF Bulk} and \eqref{DAF Surface} the
following terms must, respectively, be added to the bulk and surface
Euler-Lagrange conditions 
\[
-DIV_{X}\left(DIV_{X}\left(\frac{\partial W^{II}}{\partial\nabla F}\right)\right);\quad DIV_{X}\left(\frac{\partial W^{II}}{\partial\nabla F}\right)\cdot N
\]

\subsection{First gradient surface stress}

We now have to treat the first term in \eqref{DADF FINAL}, performing
a surface integration by parts. We obtain 
\begin{equation}
-\int_{\partial\mathcal{B}}\left(\frac{\partial W^{II}}{\partial F_{A,B}^{a}}N_{B}\right)\delta F_{A}^{a}dA.\label{Surface Stress}
\end{equation}
Recall that in dell'Isola et al. \cite{dellisolaseppechermadeo2012}
the factor 
\[
\left(\frac{\partial W^{II}}{\partial F_{A,B}^{a}}N_{B}\right)
\]
appearing in a virtual work functional of the kind given in \eqref{Surface Stress}
was called first gradient surface stress. To proceed in the calculations
we need to use some results from Gaussian differential geometry (see
e.g. Appendices for more details). The main tool we use consists in
the introduction of two projector fields $P$ and $Q$ in the neighborhood
of the surface $\partial\mathcal{B}$. The operator $P$ projects
onto its tangent plane, while $Q$ projects on the normal. The used
integration-by-parts techniques were introduced to us by Seppecher
\cite{SEPPECHER1987}. They are developed in the framework of Levi-Civita
absolute tensor calculus, however it is clear that the sources of
Berdichevsky \cite{Berdi} systematically employed these techniques.
With their help, the expression \eqref{Surface Stress} is transformed
in the following way 
\begin{align}
-\int_{\partial\mathcal{B}}\left(\frac{\partial W^{II}}{\partial F_{A,B}^{a}}N_{B}\right)\delta\chi_{,A}^{a}dA & =-\int_{\partial\mathcal{B}}\left(\frac{\partial W^{II}}{\partial F_{A,B}^{a}}N_{B}\right)\delta\chi_{,C}^{a}\delta_{A}^{C}dA=-\int_{\partial\mathcal{B}}\left(\frac{\partial W^{II}}{\partial F_{A,B}^{a}}N_{B}\right)\delta\chi_{,C}^{a}\left(Q_{A}^{C}+P_{A}^{C}\right)dA\nonumber \\
 & =-\int_{\partial\mathcal{B}}\left(\frac{\partial W^{II}}{\partial F_{A,B}^{a}}N_{B}\right)\delta\chi_{,C}^{a}Q_{D}^{C}Q_{A}^{D}dA-\int_{\partial\mathcal{B}}\left(\frac{\partial W^{II}}{\partial F_{A,B}^{a}}N_{B}\right)\delta\chi_{,C}^{a}P_{D}^{C}P_{A}^{D}dA\label{FIRSTGRADSURFACESTRESS}
\end{align}
In the following subsections, each elementary term will be discussed.

\subsection{External and contact surface double forces}

Considering that 
\[
Q_{D}^{C}:=N^{C}N_{D}
\]
the first term in equation \eqref{FIRSTGRADSURFACESTRESS} is rewritten
\[
-\int_{\partial\mathcal{B}}\left(\frac{\partial W^{II}}{\partial F_{A,B}^{a}}N_{B}\right)\delta\chi_{,C}^{a}Q_{D}^{C}Q_{A}^{D}dA=-\int_{\partial\mathcal{B}}\left(\frac{\partial W^{II}}{\partial F_{A,B}^{a}}N_{B}\right)\delta\chi_{,C}^{a}N^{C}N_{D}N^{D}N_{A}dA=-\int_{\partial\mathcal{B}}\left(\frac{\partial W^{II}}{\partial F_{A,B}^{a}}N_{B}N_{A}\right)\left(\delta\chi_{,C}^{a}N^{C}\right)dA
\]
or, in coordinate-free form, 
\begin{equation}
-\int_{\partial\mathcal{B}}\left(\frac{\partial W^{II}}{\partial\nabla F}\cdot\left(N\otimes N\right)\right)\cdot\left(\frac{\partial\delta\chi}{\partial N}\right)dA.\label{WorkDoubleForces}
\end{equation}
This last expression cannot be reduced further, and makes clear the
appearance of a new kind of boundary condition. This quantity represents
the work expended on the independent kinematical quantity 
\[
\frac{\partial\delta\chi}{\partial N}
\]
by its dual action, which is sometimes called a \emph{double force}
(see e.g Germain \cite{GERMAIN1973}), namely 
\[
\frac{\partial W^{II}}{\partial\nabla F}\cdot\left(N\otimes N\right).
\]
Actually the appearance of the work functional \eqref{WorkDoubleForces}
justifies the following statement, which fits in the spirit of Piola
\cite{Piola4} and is reaffirmed in Berdichevsky \cite{Berdi}:

\textit{Second gradient continua can sustain external surface double
forces, } \textit{i.e. external actions expending work on the virtual
normal gradient of displacement fields.}

\noindent As a consequence, in the action functional, one is allowed
to add a term of the kind: 
\[
\mathcal{A}_{S}^{II}=\int_{t_{o}}^{t_{1}}\int_{\partial\mathcal{B}}\left(-W_{S}^{II}(\chi,\frac{\partial\chi}{\partial N},X)\right)dAdt
\]
where the potential $W_{S}^{II}(\chi,\frac{\partial\chi}{\partial N},X)$
can be called \emph{surface external double potential}.

\subsection{Edge contact forces}

The term expressing the work expended on virtual displacement fields
parallel to the tangent space to $\partial\mathcal{B}$, namely 
\[
\int_{\partial\mathcal{B}}\delta\chi_{,C}^{a}P_{D}^{C}P_{A}^{D}\left(\frac{\partial W^{II}}{\partial F_{A,B}^{a}}N_{B}\right)dA
\]
can be reduced by means of an integration by parts in the submanifold
$\partial\mathcal{B}$ to

\begin{equation}
\int_{\partial\mathcal{B}}\left(\delta\chi_{,C}^{a}P_{A}^{D}\frac{\partial W^{II}}{\partial F_{A,B}^{a}}N_{B}\right)P_{D}^{C}dA=\int_{\partial\mathcal{B}}\frac{\partial}{\partial X^{C}}\left(P_{A}^{D}\frac{\partial W^{II}}{\partial F_{A,B}^{a}}N_{B}\delta\chi^{a}\right)P_{D}^{C}dA-\int_{\partial\mathcal{B}}\frac{\partial}{\partial X^{C}}\left(P_{A}^{D}\frac{\partial W^{II}}{\partial F_{A,B}^{a}}N_{B}\right)\delta\chi^{a}P_{D}^{C}dA.\label{eq:edge cont}
\end{equation}
Surface divergence theorem is then applied to the first term, resulting
in the following equality (see Appendices or dell'Isola et al. \cite{dellisolaseppechermadeo2012})
\begin{gather}
\int_{\partial\mathcal{B}}\frac{\partial}{\partial X^{C}}\left(P_{A}^{D}\frac{\partial W^{II}}{\partial F_{A,B}^{a}}N_{B}\delta\chi^{a}\right)P_{D}^{C}dA=\int_{\partial\partial\mathcal{B}}\left(\frac{\partial W^{II}}{\partial F_{A,B}^{a}}N_{B}\delta\chi^{a}\right)P_{A}^{C}\nu_{C}dL=\int_{\partial\partial\mathcal{B}}\delta\chi^{a}\left(\frac{\partial W^{II}}{\partial F_{A,B}^{a}}N_{B}\nu_{A}\right)dL\label{edge forces}
\end{gather}
When the surface $\partial\mathcal{B}$ is orientable and $C^{1}$,
the boundary $\partial\partial\mathcal{B}$ is empty. Alternatively
, if $\partial\mathcal{B}$ is piecewise $C^{1}$ then $\partial\partial\mathcal{B}$
is the union of the edges of $\partial\mathcal{B}$ and the obtained
expression represents the work expended by contact edge forces on
the virtual displacement $\delta\chi.$ To the boundary conditions
it is therefore necessary to add on $\partial\partial\mathcal{B}$
the following terms, which balance external line forces 
\[
\frac{\partial W^{II}}{\partial F_{A,B}^{i}}N_{B}\nu_{A}.
\]
Once again, the appearance of the work functional \eqref{edge forces}
justifies the following statement:\medskip{}

\begin{center}
\textbf{\textit{Second gradient continua can sustain external line
forces, }}\textbf{ }\textbf{\textit{i.e. external actions expending
work on virtual displacement fields at the edges of the boundary}}\textbf{
$\partial\mathcal{B}$. }\medskip{}

\par\end{center}

\noindent This means that, in the action functional, one is allowed
to add a term of the kind: 
\[
\mathcal{A}_{L}^{II}=\int_{t_{o}}^{t_{1}}\int_{\partial\partial\mathcal{B}}\left(-W_{L}^{II}(\chi,X)\right)dLdt
\]
where the potential $W_{L}^{II}(\chi,X)$ can be called \emph{line
external potential}.

\subsection{Contact forces depending on curvature of contact surfaces}

The second term of equation \eqref{eq:edge cont} produces a further
term to be added to surface boundary conditions, which can be interpreted
as a new kind of contact force (as it expends work on virtual displacements).
The newly (by Casal, Mindlin, Green, Rivlin and Germain) found contact
force does not \emph{obey} the so-called Cauchy postulate, as it depends
not only on the normal of Cauchy cuts but also on their curvature.
The surface boundary conditions have to be complemented by the following
terms 
\[
-DIV_{\partial\mathcal{B}}\left(P\left(\frac{\partial W^{II}}{\partial\nabla F}\cdot N\right)\right).
\]
which depend explicitly on the curvature of the surface $\partial\mathcal{B}$.

\subsection{Resumé of terms to be added to Euler-Lagrange equations for second
gradient continua}

The Euler-Lagrange conditions found for first gradient action have
to be completed by the terms listed below (see \cite{dellisolaseppecher1995,dellisolaseppecher1997}): 
\begin{itemize}
\item terms to be added to bulk equations 
\begin{equation}
DIV_{X}\left(\frac{\partial W^{II}}{\partial F}\right)-\frac{\partial W^{II}}{\partial\chi^{i}}-DIV_{X}\left(DIV_{X}\left(\frac{\partial W^{II}}{\partial\nabla F}\right)\right)\label{BULK EULER-LAGRANGE IIGRAD}
\end{equation}

\item terms to be added to surface boundary conditions 
\begin{equation}
-\frac{\partial W^{II}}{\partial F}\cdot N+DIV_{X}\left(\frac{\partial W^{II}}{\partial\nabla F}\right)\cdot N-DIV_{\partial\mathcal{B}}\left(P\left(\frac{\partial W^{II}}{\partial\nabla F}\cdot N\right)\right)\label{SurfaceForce}
\end{equation}

\item terms to be added to form new edge boundary conditions 
\begin{equation}
\frac{\partial W^{II}}{\partial\nabla F}\cdot\left(N\otimes\nu\right)-\frac{\partial W_{L}^{II}(\chi,X)}{\partial\chi}\label{LineForce}
\end{equation}

\item terms forming new surface boundary conditions (which may be called
\emph{balance of contact double forces}) 
\begin{equation}
\frac{\partial W^{II}}{\partial\nabla F}\cdot\left(N\otimes N\right)-\frac{\partial W_{S}^{II}(\chi,\frac{\partial\chi}{\partial N},X)}{\partial\left(\frac{\partial\chi}{\partial N}\right)}\label{SurfaceDoubleForce}
\end{equation}

\end{itemize}

\subsection{Objective second gradient energies}

The added term 
\[
W^{II}(\chi,F,\nabla F,X)
\]
must of course be invariant under the change of the observer in the
Eulerian configuration. The use of the Cauchy-Green deformation tensor
ensures that the deformation energy is objective (see e.g. \cite{dellisolasciarravidoli2009}).
This requirement is verified by a deformation energy having one of
the forms 
\[
\hat{W}^{II}(C,\nabla C,X);\quad\check{W}^{II}(C^{-1},\nabla C^{-1},X)
\]
It is interesting to remark that many continuum models of fiber reinforced
materials (see e.g Steigmann \cite{Steigmann1}, Atai and Steigmann
\cite{AtaiSteigmann}, Nadler and Steigmann \cite{Nadler1}, Nadler
et al. \cite{Nadler2}, Haseganu and Steigmann \cite{Haseganu}) show
some peculiarities which can be explained by the introduction of second
gradient or even higher gradient models. Therefore, in order to calculate
the partial derivatives with respect to $F$ and $\nabla F$ appearing
in the equations \eqref{BULK EULER-LAGRANGE IIGRAD}, \eqref{SurfaceForce},
\eqref{LineForce} and \eqref{SurfaceDoubleForce}, it is necessary
to calculate the derivatives listed in the following formulas (see
Appendix \ref{app:Basic-Kinematical-Formulas} for more details ). 
\begin{itemize}
\item Derivatives of $C$ and $\nabla C$ : 
\end{itemize}
\[
\begin{aligned}\frac{\partial C_{MN}}{\partial F_{P}^{i}} & =\delta_{M}^{P}F_{iN}+F_{iM}\delta_{N}^{P}\\
\frac{\partial C_{MN,O}}{\partial F_{P}^{i}} & =F_{iM,O}\delta_{P}^{N}+F_{iN,O}\delta_{P}^{M}\\
\frac{\partial C_{MN,O}}{\partial F_{P,Q}^{i}} & =\delta_{M}^{P}\delta_{O}^{Q}F_{Ni}+\delta_{N}^{P}\delta_{O}^{Q}F_{Mi}
\end{aligned}
\]

\begin{itemize}
\item Derivatives of $C^{-1}$ and $\nabla C^{-1}$ : 
\end{itemize}
\[
\begin{aligned}\frac{\partial C_{MN}^{-1}}{\partial F_{P}^{i}} & =-\left(F^{-1}\right)_{Mi}\left(F^{-1}\right)_{a}^{P}\left(F^{-1}\right)_{N}^{a}-\left(F^{-1}\right)_{Ni}\left(F^{-1}\right)^{bP}\left(F^{-1}\right)_{bM}\\
\frac{\partial C_{MN,O}^{-1}}{\partial F_{P}^{l}} & =-\left(F^{-1}\right)^{aP}\left(\left(F^{-1}\right)_{Nl}\left(F^{-1}\right)_{Ma,O}+\left(F^{-1}\right)_{Ml}\left(F^{-1}\right)_{aN,O}\right)\\
\frac{\partial C_{MN,O}^{-1}}{\partial F_{P,Q}^{i}} & =-\left[\left(F^{-1}\right)_{Mi}\left(F^{-1}\right)^{aP}\left(F^{-1}\right)_{aN}+\left(F^{-1}\right)_{Ni}\left(F^{-1}\right)^{bP}\left(F^{-1}\right)_{bM}\right]\delta_{Q}^{O}
\end{aligned}
\]

\subsection{Capillary fluids}

In Poisson \cite{poisson3} pages 5-6: (translated by the authors)
one finds the following statements about the region of a fluid in
which a phase transition occurs (page 5)

\emph{``But Laplace omitted, in his calculations, a physical circumstance
whose consideration is essential: I refer to the rapid variation of
density which the liquid experiences in proximity of its free surface
and of the tube wall, {[}variation{]} without which the capillary
phenomena could not occur {[}....{]} Actually, in an equilibrium state,
each layer infinitely thin of a liquid is compressed equally on both
of its faces by the repulsive actions of all close molecules diminished
by their attractive force {[}....{]} and its level of condensation
is determined by the magnitude of the compressive force. At a sensible
distance from the surface of the liquid the aforementioned force is
exerted by a liquid layer adjacent to the infinitely thin layer, whose
thickness is complete and everywhere constant, i.e. equal to the radius
of activity of fluid molecules; and for this reason the internal density
of the liquid is also constant {[}...{]} But when this distance is
less than the radius of molecular activity the thickness of the layer
under the layer which we are considering is also smaller than this
radius: the compressive force which is exerted by the said upper layer
is therefore decreasing very rapidly} \emph{with the distance from
the surface and vanishes at the surface itself, where the infinitesimal
thin layer is compressed only by the atmospheric pressure. Consequently,
the condensation of the liquid is also decreasing, following an unknown
law, when one is approaching its free surface and its density is very
different in that surface and at a depth which exceeds by a small
amount the activity radius of its molecules, which is sufficient for
having this density to be equal to the internal density of the liquid.
Now it will be proven in the first chapter of this work that if one
neglects this rapid variation of density in the thickness of the interfacial
layer}%
\footnote{This thickness must have a finite value but this value must be undetectable
not sensible, because of the hypothesis which was accepted about the
extension of molecular activity. This is confirmed by the experience
made by M.Gay-Lussac.%
}\emph{ then the capillary surface should result to be plane and horizontal
and one could not observe neither elevation nor lowering of the liquid
level.{[}...{]}''}

Therefore we can conclude that already Poisson wanted, with some assumptions
which probably need to be clarified, to model the interfacial layer
as a thin but three-dimensional layer. It is interesting to remark
that it is only because of the development of the ideas by Piola (ideas
which Poisson violently criticized) that the modern theory of capillary
fluids managed to give a precise meaning to the Poisson's intuitions.
What Poisson calls \emph{an unknown law} is now explicitly determined
by using second gradient continua (see e.g. \cite{CASAL1961,SEPPECHER1988}).

In the spirit of Piola's works, we now consider the most simple class
of second gradient continua, i.e. capillary fluids. We recall here
that capillary fluids are continua whose Eulerian volumetric deformation
energy density depends both on their Eulerian mass density $\rho$
and on its gradient $\nabla\rho$. For capillary fluids an additive
extra term in the part of action related to deformation energy has
to be considered: 
\[
\mathcal{A}^{\mathrm{cap}}=\int_{\mathcal{E}}\hat{W}^{\mathrm{cap}}\left(\rho,\nabla\rho\right)dv=\int_{\mathcal{B}}J\hat{W}^{\mathrm{cap}}\left(\left(\rho\right)^{\overrightarrow{\left(\mathcal{B}\right)}},\left(\nabla\rho\right)^{\overrightarrow{\left(\mathcal{B}\right)}}\right)dV
\]
The notations $\left(\cdot\right)^{\overrightarrow{\left(\mathcal{B}\right)}}$
and $\left(\cdot\right)^{\overrightarrow{\left(\mathcal{E}\right)}}$
introduced in the Appendix A, will be omitted occasionally for the
sake of readability. Obviously the dependence of $\hat{W}^{\mathrm{cap}}$
on $\nabla\rho$ must be objective. Therefore we must have (see e.g.
Ball \cite{ball1976}) 
\begin{equation}
\hat{W}^{\mathrm{cap}}\left(\rho,\nabla\rho\right)=\check{W}^{\mathrm{cap}}\left(\rho,\beta\right)\label{EnergyCapillary}
\end{equation}
where we introduced the scalar 
\[
\beta:=\nabla\rho\cdot\nabla\rho.
\]
A particular case of the energy \eqref{EnergyCapillary} is given
by the one discussed by Cahn and Hilliard 
\[
\check{W}^{\mathrm{cap}}\left(\rho,\beta\right)=\frac{1}{2}\lambda\left(\rho\right)\beta=\frac{1}{2}\lambda\left(\rho\right)\left(\nabla\rho\cdot\nabla\rho\right)
\]
where the function $\lambda\left(\rho\right)$ has been often considered
to be constant.

\subsubsection{Lagrangian expression for the deformation energy of capillary fluids}

It is therefore needed to calculate the following first variation
\[
\delta\mathcal{A}^{\mathrm{cap}}=\delta\left(\int_{\mathcal{B}}J\check{W}^{\mathrm{cap}}\left(\left(\rho\right)^{\overrightarrow{\left(\mathcal{B}\right)}},\left(\beta\right)^{\overrightarrow{\left(\mathcal{B}\right)}}\right)dV\right).
\]
Once we have defined (with an abuse of notation) 
\begin{equation}
W^{\mathrm{cap}}(F,\nabla F):=J\check{W}^{\mathrm{cap}}\left(\left(\rho\right)^{\overrightarrow{\left(\mathcal{B}\right)}},\left(\beta\right)^{\overrightarrow{\left(\mathcal{B}\right)}}\right)=\frac{\rho_{0}}{\left(\rho\right)^{\overrightarrow{\left(\mathcal{B}\right)}}}\check{W}^{\mathrm{cap}}\label{GeneralEnergyCapillary}
\end{equation}
it is clear that 
\begin{alignat*}{1}
\delta\left(J\check{W}^{\mathrm{cap}}\right) & =\left(\check{W}^{\mathrm{cap}}\delta J+J\frac{\partial\check{W}^{\mathrm{cap}}}{\partial\left(\rho\right)^{\overrightarrow{\left(\mathcal{B}\right)}}}\delta\left(\rho\right)^{\overrightarrow{\left(\mathcal{B}\right)}}+J\frac{\partial\check{W}^{\mathrm{cap}}}{\partial\left(\beta\right)^{\overrightarrow{\left(\mathcal{B}\right)}}}\delta\left(\beta\right)^{\overrightarrow{\left(\mathcal{B}\right)}}\right)\\
 & =\left(\check{W}^{\mathrm{cap}}\frac{\partial J}{\partial F}\delta F+J\frac{\partial\check{W}^{\mathrm{cap}}}{\partial\left(\rho\right)^{\overrightarrow{\left(\mathcal{B}\right)}}}\frac{\partial\left(\rho\right)^{\overrightarrow{\left(\mathcal{B}\right)}}}{\partial F}\delta F\right)+\frac{\partial\check{W}^{\mathrm{cap}}}{\partial\left(\beta\right)^{\overrightarrow{\left(\mathcal{B}\right)}}}J\left(\frac{\partial\left(\beta\right)^{\overrightarrow{\left(\mathcal{B}\right)}}}{\partial F}\delta F+\frac{\partial\left(\beta\right)^{\overrightarrow{\left(\mathcal{B}\right)}}}{\partial\nabla F}\delta\nabla F\right)
\end{alignat*}
As a consequence (with another abuse of notation) we have

\begin{align}
\frac{\partial W^{\mathrm{cap}}}{\partial F}= & \check{W}^{\mathrm{cap}}\frac{\partial J}{\partial F}+J\frac{\partial\check{W}^{\mathrm{cap}}}{\partial\rho}\frac{\partial\rho}{\partial F}+J\frac{\partial\check{W}^{\mathrm{cap}}}{\partial\beta}\frac{\partial\beta}{\partial F}\label{DEWCAPDEF}\\
\frac{\partial W^{\mathrm{cap}}}{\partial\nabla F}= & J\frac{\partial\check{W}^{\mathrm{cap}}}{\partial\beta}\frac{\partial\beta}{\partial\nabla F}\label{DEWCAPDEGRADF}
\end{align}

\subsubsection{Eulerian balance equations for capillary fluids}

Following the original methods introduced by Piola, after having applied
the principle of least action or the principle of virtual work in
the Lagrangian description, we must transform the obtained stationarity
conditions into some other conditions which are valid in the Eulerian
description. As previously seen, in Lagrangian description the balance
equations for capillary fluids read 
\begin{equation}
-\frac{\partial}{\partial t}\left(\rho_{0}v_{i}\right)+DIV_{X}\left(\frac{\partial W^{\mathrm{eul}}}{\partial F}+\frac{\partial W^{\mathrm{cap}}}{\partial F}\right)-DIV_{X}\left(DIV_{X}\left(\frac{\partial W^{\mathrm{cap}}}{\partial\nabla F}\right)\right)=0\label{CapillaryBalanceLagrange}
\end{equation}
where $W^{\mathrm{eul}}$ and $W^{\mathrm{cap}}$ were defined, respectively,
in \eqref{EnergyEulerFluids} and \eqref{GeneralEnergyCapillary}.
The terms in \eqref{CapillaryBalanceLagrange}, which are specific
to capillary fluids, must therefore be estimated. Starting from equation
\eqref{DEWCAPDEF} and using the following result (calculated in \eqref{app:dbetadef}
and \eqref{app:derhodf})

\[
\frac{\partial J}{\partial F_{M}^{i}}=J\left(F^{-1}\right)_{i}^{M},\quad\frac{\partial\rho}{\partial F_{M}^{i}}=-\rho\left(F^{-1}\right)_{i}^{M},\quad\frac{\partial\beta}{\partial F_{M}^{i}}=-2g^{ab}\left(\rho_{,a}\rho_{,i}\left(F^{-1}\right)_{b}^{M}+\rho_{,a}\left(\rho\left(F^{-1}\right)_{i}^{M}\right)_{,b}\right)
\]
we obtain (the notation $\left(\cdot\right)^{\overrightarrow{\left(\mathcal{B}\right)}}$
has been dropped to yield more readable formulas),

\begin{align}
\frac{\partial W^{\mathrm{cap}}}{\partial F_{M}^{i}} & =-\mathcal{P}^{\mathrm{cap}}J\left(F^{-1}\right)_{i}^{M}-2\frac{\partial\check{W}^{\mathrm{cap}}}{\partial\beta}J\left(g^{ab}\rho_{,a}\rho_{,i}\left(F^{-1}\right)_{b}^{M}+\beta\left(F^{-1}\right)_{i}^{M}+g^{ab}\rho_{,a}\rho\left(F^{-1}\right)_{i,b}^{M}\right)\label{DEWDEF}
\end{align}
where we have introduced 
\[
\mathcal{P}^{\mathrm{cap}}:=\rho\frac{\partial\check{W}^{\mathrm{cap}}}{\partial\rho}-\check{W}^{\mathrm{cap}}
\]

\subsubsection{Piola stress decomposition}

In the remaining part of the paper, different Piola stress tensors
will be considered. Therefore, and in order to avoid any misunderstanding,
some time will be devoted to properly define these different stress
tensors. This discussion is specific to higher-order continua, since
for the first gradient continuum these different tensors are either
identical or null. As a starting point we define the\emph{ bulk Piola}
stress for capillary fluids by 
\begin{equation}
\mathsf{\mathbb{P}}_{i}^{M}:=\frac{\partial W^{\mathrm{eul}}}{\partial F_{M}^{i}}+\frac{\partial W^{\mathrm{cap}}}{\partial F_{M}^{i}}-\frac{\partial}{\partial X^{A}}\left(\frac{\partial W^{\mathrm{cap}}}{\partial F_{M,A}^{i}}\right)\label{piolastress}
\end{equation}
as the quantity that appears in the Lagrangian balance equation 
\[
-\frac{\partial}{\partial t}\left(\rho_{0}v_{i}\right)+\frac{\partial\mathsf{\mathbb{P}}_{i}^{A}}{\partial X^{A}}-\frac{\partial U^{\mathrm{ext}}}{\partial\chi^{i}}=0
\]
This\emph{ }tensor is an\emph{ effective tensor }(in the following
effective tensor are written using blackboard fonts) since\emph{ }it
is composed of tensors of different order, as

\[
\mathsf{\mathbb{P}}_{i}^{M}:=\mathrm{P}_{i}^{M}+\frac{\partial}{\partial X^{A}}\left(\mathsf{H}_{i}^{MA}\right),
\]
where $\mathrm{P}_{i}^{M}$is the classical Piola stress, and $\mathsf{H}_{i}^{MA}$
is third-order \emph{Hyper Piola} stress defined as

\[
\mathsf{H}_{i}^{MN}:=\frac{\partial W}{\partial F_{M,N}^{i}}.
\]
It is worth noting that for capillary fluids, the classical Piola
stress can be decomposed as

\[
\mathrm{P}_{i}^{M}:=(\mathrm{P}^{\mathrm{eul}})_{i}^{M}+(\mathrm{P}^{\mathrm{cal}})_{i}^{M}
\]
Hence, another \emph{effective} tensor can be defined 
\[
\mathsf{(\mathbb{P}^{\mathrm{cal}})}_{i}^{M}:=(\mathrm{P}^{\mathrm{cal}})_{i}^{M}+\frac{\partial}{\partial X^{A}}\left(\mathsf{H}_{i}^{MA}\right)
\]
resulting in the following additive decomposition of the\emph{ bulk
Piola} stress

\[
\mathsf{\mathbb{P}}_{i}^{M}:=(\mathrm{P}^{\mathrm{eul}})_{i}^{M}+\mathsf{(\mathbb{P}^{\mathrm{cal}})}_{i}^{M}
\]

\subsubsection{Piola stress for capillary fluids}

Now we will effectively compute the effective\emph{ bulk Piola tensor.
}To that aim, we start by calculating the relevant term by using \eqref{DEWCAPDEGRADF}
and \eqref{DEBETADEGRADF} thus, 
\[
\frac{\partial\beta}{\partial F_{M,N}^{i}}=-2g^{ab}\rho\rho_{,a}\left(F^{-1}\right)_{i}^{M}\left(F^{-1}\right)_{b}^{N},
\]
\begin{align*}
-\frac{\partial}{\partial X^{A}}\left(\frac{\partial\check{W}^{\mathrm{cap}}}{\partial F_{M,A}^{i}}\right) & =-\frac{\partial}{\partial X^{A}}\left(J\frac{\partial\check{W}^{\mathrm{cap}}}{\partial\beta}\frac{\partial\beta}{\partial F_{M,A}^{i}}\right)=\frac{\partial}{\partial X^{A}}\left(\rho_{0}2g^{ab}\rho_{,a}\frac{\partial\check{W}^{\mathrm{cap}}}{\partial\beta}\left(F^{-1}\right)_{i}^{M}\left(F^{-1}\right)_{b}^{A}\right)\\
 & =\frac{\partial}{\partial X^{A}}\left(\rho_{0}2g^{ab}\rho_{,a}\frac{\partial\check{W}^{\mathrm{cap}}}{\partial\beta}\left(F^{-1}\right)_{i}^{M}\right)\left(F^{-1}\right)_{b}^{A}+\rho_{0}2g^{ab}\rho_{,a}\frac{\partial\check{W}^{\mathrm{cap}}}{\partial\beta}\left(F^{-1}\right)_{i}^{M}\frac{\partial}{\partial X^{A}}\left(\left(F^{-1}\right)_{b}^{A}\right)\\
 & =\frac{\partial}{\partial x^{b}}\left(\rho_{0}2g^{ab}\rho_{,a}\frac{\partial\check{W}^{\mathrm{cap}}}{\partial\beta}\left(F^{-1}\right)_{i}^{M}\right)+\rho_{0}2g^{ab}\rho_{,a}\frac{\partial\check{W}^{\mathrm{cap}}}{\partial\beta}\left(F^{-1}\right)_{i}^{M}\left(F^{-1}\right)_{b,A}^{A}.
\end{align*}
Now we use \eqref{CorollaryPiolaCondition}, i.e. 
\[
\left(F^{-1}\right)_{i,A}^{A}=\frac{\rho_{,A}}{\rho}\left(F^{-1}\right)_{i}^{A}=\frac{\rho_{,i}}{\rho}
\]
to derive 
\begin{align}
-\frac{\partial}{\partial X^{A}}\left(\frac{\partial W^{\mathrm{cap}}}{\partial F_{M,A}^{i}}\right) & =\frac{\partial}{\partial x^{b}}\left(2\frac{\partial\check{W}^{\mathrm{cap}}}{\partial\beta}\rho_{0}g^{ab}\rho_{,a}\right)\left(F^{-1}\right)_{i}^{M}+2\frac{\partial\check{W}^{\mathrm{cap}}}{\partial\beta}\rho_{0}g^{ab}\rho_{,a}\left(F^{-1}\right)_{i,b}^{M}+2\beta\frac{\partial\check{W}^{\mathrm{cap}}}{\partial\beta}J\left(F^{-1}\right)_{i}^{M}.\label{DIVDEWCAPGRADF}
\end{align}
Using \eqref{DEWDEF} and \eqref{DIVDEWCAPGRADF} in \eqref{piolastress}
we obtain 
\[
\mathsf{\mathbb{P}}_{i}^{M}=\left(-\left(p\left(\rho\right)+\mathcal{P}^{\mathrm{cap}}\right)+\rho\frac{\partial}{\partial x^{b}}\left(2\frac{\partial\check{W}^{\mathrm{cap}}}{\partial\beta}g^{ab}\rho_{,a}\right)\right)J\left(F^{-1}\right)_{i}^{M}-2\frac{\partial\check{W}^{\mathrm{cap}}}{\partial\beta}g^{ab}\rho_{,a}\rho_{,i}J\left(F^{-1}\right)_{b}^{M},
\]
where we have used 
\[
\frac{\partial W^{\mathrm{eul}}}{\partial F_{M}^{i}}=-Jp\left(\rho\right)\left(F^{-1}\right)_{i}^{M}.
\]

\subsubsection{Cauchy stress for capillary fluids}

As for the effective bulk Piola stress, we define the effective bulk
Cauchy stress as the quantity that appears in the Eulerian balance
equation

\[
-\rho\left(\frac{\partial v_{i}^{\overrightarrow{(\mathcal{E})}}}{\partial t}+\frac{\partial v_{i}^{\overrightarrow{(\mathcal{E})}}}{\partial x^{a}}\left(v^{a}\right)^{\overrightarrow{(\mathcal{E})}}\right)-\frac{\partial}{\partial x^{b}}\left(\mathsf{\mathbb{T}}_{i}^{b}\right)-\rho\left(\frac{\partial\left(U^{exp}/\rho_{0}\right)}{\partial\chi^{i}}\right)^{\overrightarrow{(\mathcal{E})}}=0.
\]
This\emph{ effective tensor }can be decomposed as

\[
\mathsf{\mathbb{T}}_{i}^{j}:=\mathrm{T}_{i}^{j}+\frac{\partial}{\partial x^{A}}\left(\mathsf{S}_{i}^{ja}\right),
\]
where $\mathrm{T}_{i}^{j}$is the second-order capillary Cauchy stress,
and $\mathsf{S}_{i}^{jk}$ is the third-order \emph{capillary Hyper
Cauchy} stress. As previously explained, the second-order Cauchy stress
can be decomposed as

\[
\mathrm{T}_{i}^{j}:=(\mathrm{T}^{\mathrm{eul}})_{i}^{j}+(\mathrm{T}^{\mathrm{cap}})_{i}^{j}
\]
Hence, another \emph{effective} tensor can be defined 
\[
\mathsf{(\mathbb{T}^{\mathrm{cap}})}_{i}^{j}:=(\mathrm{T}^{\mathrm{cap}})_{i}^{j}+\frac{\partial}{\partial x^{a}}\left(\mathsf{S}_{i}^{ja}\right),
\]
resulting in the following additive decomposition of the following\emph{
bulk Cauchy} stress:

\[
\mathsf{\mathbb{T}}_{i}^{j}:=(\mathrm{T}^{\mathrm{eul}})_{i}^{j}+\mathsf{(\mathbb{T}^{\mathrm{cap}})}_{i}^{j}.
\]
Let us now return to the explicit determination of $\mathsf{\mathbb{T}}_{i}^{j}$.
By recalling (see Appendix A) the Piola transformation of tensors
from the Lagrangian to the Eulerian description, i.e. 
\[
\mathsf{\mathbb{T}}_{i}^{j}=J^{-1}\left(\mathsf{\mathbb{P}}_{i}^{A}F_{A}^{j}\right)^{\overrightarrow{(\mathcal{E})}},
\]
the \emph{bulk Cauchy} stress tensor for capillary fluids is obtained
as 
\[
\mathsf{\mathbb{T}}_{i}^{j}=\left(-\left(p\left(\rho\right)+\mathcal{P}^{\mathrm{cap}}\right)+\rho\frac{\partial}{\partial x^{b}}\left(2\frac{\partial\check{W}^{\mathrm{cap}}}{\partial\beta}g^{ab}\rho_{,a}\right)\right)\delta_{i}^{j}-2\frac{\partial\check{W}^{\mathrm{cap}}}{\partial\beta}g^{aj}\rho_{,a}\rho_{,i}.
\]
In the case of Cahn-Hilliard fluids with a constant $\lambda$ we
have 
\[
2\frac{\partial\check{W}^{\mathrm{cap}}}{\partial\beta}=\lambda,\qquad\check{W}^{\mathrm{cap}}=-\mathcal{P}^{\mathrm{cap}}=\frac{\lambda}{2}g^{ab}\rho_{,a}\rho_{,b},
\]
so that 
\[
\mathsf{\mathbb{T}}_{i}^{j}=\left(-p\left(\rho\right)+\frac{\lambda}{2}g^{ab}\rho_{,a}\rho_{,b}+\rho\frac{\partial}{\partial x^{b}}\left(\lambda g^{ab}\rho_{,a}\right)\right)\delta_{i}^{j}-\lambda g^{aj}\rho_{,a}\rho_{,i},
\]
which is exactly the result found in the literature (see Seppecher
\cite{SEPPECHER1989}\ or Casal and Gouin \cite{Casal-Gouin1985,CASALGOUIN1988}).
Let us now develop the Eulerian divergence of the effective capillary
Cauchy tensor: 
\begin{align*}
\frac{\partial}{\partial x^{c}}\left(\mathsf{\mathbb{T}}_{i}^{c}\right) & =\frac{\partial}{\partial x^{c}}\left(\left(-p\left(\rho\right)+\frac{\lambda}{2}g^{ab}\rho_{,a}\rho_{,b}+\rho\frac{\partial}{\partial x^{b}}\left(\lambda g^{ab}\rho_{,a}\right)\right)\delta_{i}^{c}-\lambda g^{ac}\rho_{,a}\rho_{,i}\right)\\
 & =-\frac{\partial}{\partial x^{i}}p\left(\rho\right)+\lambda g^{ab}\rho_{,a}\rho_{,bi}+\frac{\partial}{\partial x^{i}}\left(\rho\lambda g^{ab}\rho_{,ab}\right)-\lambda g^{ac}\rho_{,ac}\rho_{,i}-\lambda g^{ac}\rho_{,a}\rho_{,ic}\\
 & =-\frac{\partial}{\partial x^{i}}p\left(\rho\right)+\lambda\rho\frac{\partial}{\partial x^{i}}\left(g^{ab}\rho_{,ab}\right).
\end{align*}
In conclusion the Eulerian balance equation for Cahn-Hilliard fluids
is: 
\[
-\rho\left(\frac{\partial v_{i}^{\overrightarrow{(\mathcal{E})}}}{\partial t}+\frac{\partial v_{i}^{\overrightarrow{(\mathcal{E})}}}{\partial x^{a}}\left(v^{a}\right)^{\overrightarrow{(\mathcal{E})}}\right)-\frac{\partial}{\partial x^{i}}p(\rho)+\lambda\rho\frac{\partial}{\partial x^{i}}\left(g^{ab}\rho_{,ab}\right)-\rho\left(\frac{\partial\left(U^{exp}/\rho_{0}\right)}{\partial\chi^{i}}\right)^{\overrightarrow{(\mathcal{E})}}=0.
\]
To complete the description of the model, the associated boundary
conditions have to be supplied.

\subsubsection{Boundary terms}

In the particular case of capillary fluids the \emph{Hyper Piola}
tensor has the following explicit expression:

\[
\mathsf{H}_{i}^{MN}=\frac{\partial W^{\mathrm{cap}}}{\partial\beta}\frac{\partial\beta}{\partial F_{M,N}^{i}}=-\lambda\rho_{0}\rho_{,a}g^{ab}\left(F^{-1}\right)_{b}^{M}\left(F^{-1}\right)_{i}^{N}.
\]
Its Eulerian equivalent is the following \emph{Hyper Cauchy} tensor

\[
\mathsf{S}_{i}^{jk}=-J^{-1}H_{i}^{AB}F_{B}^{j}F_{A}^{k}=-J^{-1}\rho_{0}g^{ab}\rho_{,a}\lambda\left(F^{-1}\right)_{i}^{B}\left(F^{-1}\right)_{b}^{A}F_{B}^{j}F_{A}^{k}=-\lambda\rho g^{ak}\rho_{,a}\delta_{i}^{j}.
\]

\paragraph{Double force }

The expression of contact double force will be discussed first. In
the absence of surface external double force, the boundary conditions
read

\[
\frac{\partial W^{\mathrm{cap}}}{\partial\nabla F}\cdot\left(N\otimes N\right)=0
\]
or, in components

\[
\frac{\partial W^{\mathrm{cap}}}{\partial F_{A,B}^{i}}N_{A}N_{B}=-\lambda\rho_{0}\rho_{,a}g^{ab}\left(F^{-1}\right)_{b}^{A}\left(F^{-1}\right)_{i}^{B}N_{A}N_{B}=0
\]
Using the Piola transformation for normals \eqref{TRANSFORMATIONOFNORMALS},
the former expression is rewritten

\[
-\lambda\rho_{0}\rho_{,a}g^{ab}\left(F^{-1}\right)_{b}^{M}\left(F^{-1}\right)_{i}^{N}J^{-1}F_{M}^{c}n_{c}J^{-1}F_{N}^{e}n_{e}=0,
\]
Hence, for line forces \eqref{LineForce} we obtain

\[
-J^{-1}\lambda\rho\rho_{,a}g^{ab}n_{b}\nu_{i}=0.
\]

\paragraph{Force}

In the absence of external force, the new boundary conditions read

\[
\mathsf{\mathbb{P}}_{i}^{A}N_{A}+\frac{\partial}{\partial X^{E}}\left(P_{C}^{D}\left(\mathsf{H}_{i}^{BC}N_{B}\right)\right)P_{D}^{E}=0,
\]
or, using the Piola transformation, in Eulerian form 
\[
\mathsf{\mathbb{T}}_{i}^{a}n_{a}+\frac{\partial}{\partial x^{e}}\left(P_{c}^{d}\left(\mathrm{S}_{i}^{bc}n_{b}\right)\right)P_{d}^{e}=0.
\]
The first term will first be considered. This term can be expanded
as

\[
\mathsf{\mathbb{T}}_{i}^{a}n_{a}=\left[\left(-p\left(\rho\right)+\frac{\lambda}{2}g^{bc}\rho_{,b}\rho_{,c}+\rho\frac{\partial}{\partial x^{c}}\left(\lambda g^{bc}\rho_{,b}\right)\right)\delta_{i}^{a}-\lambda g^{ba}\rho_{,b}\rho_{,i}\right]n_{a}
\]

\[
=\left(-p\left(\rho\right)+\frac{\lambda}{2}g^{bc}\rho_{,b}\rho_{,c}+\rho\frac{\partial}{\partial x^{c}}\left(\lambda g^{bc}\rho_{,b}\right)\right)n_{i}-\lambda g^{ab}\rho_{,b}\rho_{,i}n_{a}
\]

\[
=\left(-p\left(\rho\right)+\frac{\lambda}{2}g^{bc}\rho_{,b}\rho_{,c}+\rho\frac{\partial}{\partial x^{c}}\left(\lambda g^{bc}\rho_{,b}\right)\right)n_{i}-\lambda n^{b}\rho_{,b}\rho_{,i}
\]

\[
=\left(-p\left(\rho\right)+\frac{\lambda}{2}g^{bc}\rho_{,b}\rho_{,c}+\rho\frac{\partial}{\partial x^{c}}\left(\lambda g^{bc}\rho_{,b}\right)-\lambda n^{i}\rho_{,i}n^{b}\rho_{,b}\right)n_{i}.
\]
It remains now to consider the last part of the boundary conditions,
i.e.

\[
\frac{\partial}{\partial x^{d}}\left(P_{b}^{c}\left(\mathrm{S}_{i}^{ab}n_{a}\right)\right)P_{c}^{d}.
\]
This computation is a bit more tricky. In order to easily derive the
expression, the following identities will be established: 
\begin{eqnarray}
Q\cdot(v\otimes n) & = & (v.n)Q\label{eq:IdFir}\\
P\cdot(v\otimes n) & = & (v.n)P\nonumber 
\end{eqnarray}
Their demonstration is straightforward: 
\begin{eqnarray*}
Q_{a}^{i}v^{a}n_{j} & = & n^{i}n_{a}v^{a}n_{j}=n_{a}v^{a}n^{i}n_{j}=Q_{j}^{i}v^{a}n_{a},\\
P_{a}^{i}v^{a}n_{j} & = & (\delta_{a}^{i}-Q_{a}^{i})v^{a}n_{j}=(\delta_{j}^{i}\delta_{a}^{j}v^{a}n_{j})+Q_{j}^{i}v^{a}n_{a}\\
 & = & \delta_{j}^{i}v^{a}n_{a}+Q_{j}^{i}v^{a}n_{a}=(\delta_{j}^{i}+Q_{j}^{i})v^{a}n_{a}\\
 & = & P_{j}^{i}v^{a}n_{a}
\end{eqnarray*}
Now, using the definition of the hyperstress for capillary fluids,
we obtain 
\[
\mathrm{S}_{i}^{aj}n_{a}=-\lambda\rho\rho^{j}n_{i}.
\]
In the following, the factor $-\lambda\rho$ will dropped and only
added at the end. Using the identity \eqref{eq:IdFir} we have the
first transformation relation 
\[
P_{a}^{i}\left(\rho^{a}n_{j}\right)=\rho^{a}n_{a}P_{j}^{i}.
\]
Therefore, 
\[
\nabla_{k}^{S}\left(P_{a}^{i}\left(\rho^{a}n_{j}\right)\right)=\nabla_{k}^{S}\left(\rho^{a}n_{a}P_{j}^{i}\right)=\nabla_{k}^{S}\left(\rho^{a}n_{a}\right)P_{j}^{i}+\rho^{a}n_{a}\nabla_{k}^{S}\left(P_{j}^{i}\right),
\]
where $\nabla_{k}^{S}:=P_{k}^{a}\frac{\partial}{\partial x_{a}}$
denotes the surface (tangential) gradient. Let us now compute the
surface gradient of the projection operator $P$, 
\begin{eqnarray*}
\nabla_{k}^{S}\left(P_{j}^{i}\right) & = & \nabla_{k}^{S}\delta_{j}^{i}-\nabla_{k}^{S}(n^{i}n_{j})=-(\nabla_{k}^{S}(n^{i})n_{j}+n^{i}\nabla_{k}^{S}(n_{j}))\\
 & = & L_{k}^{i}n_{j}+n^{i}L_{kj}
\end{eqnarray*}
where $L_{ij}:=-P_{i}^{a}n_{aj}$ is the Weingarten curvature tensor.
Therefore, it follows that 
\[
\nabla_{k}^{S}\left(P_{a}^{i}\left(\rho^{a}n_{j}\right)\right)=\nabla_{k}^{S}\left(\rho^{a}n_{a}\right)P_{j}^{i}+\rho^{a}n_{a}(L_{k}^{i}n_{j}+n^{i}L_{kj}).
\]
To obtain the surface divergence it remains to multiply the previous
result by $\delta_{k}^{i}$ 
\[
\nabla_{i}^{S}\left(P_{a}^{i}\left(\rho^{a}n_{j}\right)\right)=\nabla_{i}^{S}\left(\rho^{a}n_{a}\right)P_{j}^{i}+\rho^{a}n_{a}(L_{i}^{i}n_{j}+n^{i}L_{ij}).
\]
This expression can be simplified, using 
\[
\nabla_{i}^{S}P_{j}^{i}=P_{i}^{a}\frac{\partial}{\partial x_{a}}P_{j}^{i}=P_{i}^{a}P_{j}^{i}\frac{\partial}{\partial x_{a}}=\nabla_{j}^{S},
\]
\[
n^{i}L_{ij}=n^{i}P_{i}^{a}n_{aj}=0,
\]
and 
\[
2H:=L_{i}^{i},
\]
where $H$ is the surface mean curvature. Therefore, at the end of
the journey, 
\[
\nabla_{i}^{S}\left(P_{a}^{i}\left(\rho^{a}n_{j}\right)\right)=\nabla_{j}^{S}\left(\rho^{a}n_{a}\right)+2\rho^{a}n_{a}Hn_{j}.
\]
Once the two parts added, we obtain

\[
\left(-p\left(\rho\right)+\frac{\lambda}{2}g^{ab}\rho_{,a}\rho_{,b}+\rho\frac{\partial}{\partial x^{b}}\left(\lambda g^{ab}\rho_{,a}\right)-\lambda n^{i}\rho_{,i}n^{a}\rho_{,a}+2\rho^{a}n_{a}H\right)n_{i}+\nabla_{i}^{S}\left(\rho^{a}n_{a}\right)=0,
\]
or

\[
-p^{*}n_{i}+\nabla_{i}^{S}\left(\rho^{a}n_{a}\right)=0,
\]
in which

\[
p^{*}=\left(p\left(\rho\right)-\frac{\lambda}{2}g^{ab}\rho_{,a}\rho_{,b}-\rho\frac{\partial}{\partial x^{b}}\left(\lambda g^{ab}\rho_{,a}\right)+\lambda n^{i}\rho_{,i}n^{a}\rho_{,a}+2\rho^{a}n_{a}H\right).
\]
This is exactly the result found in \cite{SEPPECHER1989,SEPPECHER1993,Casal-Gouin1985,CASALGOUIN1988}.

\subsubsection{Bernoulli Law for capillary fluids}

The results in the previous sections imply that for capillary fluids
the following Eulerian Balance of force holds (see also \cite{CASAL1972,Casal-Gouin1985})
\[
-\rho\left(\frac{\partial v_{i}^{\overrightarrow{(\mathcal{E})}}}{\partial t}+\frac{\partial v_{i}^{\overrightarrow{(\mathcal{E})}}}{\partial x^{a}}\left(v^{a}\right)^{\overrightarrow{(\mathcal{E})}}\right)-\frac{\partial}{\partial x^{i}}\left(p(\rho)\right)+\frac{\partial}{\partial x^{b}}\mathsf{(\mathbb{T}^{\mathrm{cap}})}_{i}^{b}-\rho\left(\frac{\partial U/\rho_{0}}{\partial\chi^{i}}\right)^{\overrightarrow{(\mathcal{E})}}=0,
\]
where we have introduced the constitutive equations 
\[
\mathsf{(\mathbb{T}^{\mathrm{cap}})}_{i}^{b}=\left(-\mathcal{P}^{\mathrm{cap}}\left(\rho,\beta\right)+\rho\frac{\partial}{\partial x^{b}}\left(2\frac{\partial\check{W}^{\mathrm{cap}}}{\partial\beta}g^{ab}\rho_{,a}\right)\right)\delta_{i}^{j}-2\frac{\partial\check{W}^{\mathrm{cap}}}{\partial\beta}g^{aj}\rho_{,a}\rho_{,i},
\]
\[
\begin{array}{cc}
-\mathcal{P}^{\mathrm{cap}}:=\check{W}^{\mathrm{cap}}-\rho\frac{\partial\check{W}^{\mathrm{cap}}}{\partial\rho}\quad;\quad & p(\rho):=\rho^{2}\frac{\partial\left(\Psi/\rho_{0}\right)}{\partial\rho}\end{array}.
\]
If the last relationship is invertible one can express the density
as a function $\hat{\rho}$ of the pressure and introduce the function
\[
Q(p)=\int\frac{1}{\hat{\rho}(p)}dp,
\]
which has the remarkable property 
\[
\frac{\partial Q(p)}{\partial x^{i}}=\frac{1}{\hat{\rho}(p)}\frac{\partial p}{\partial x^{i}}.
\]
As a consequence, once divided by $\rho$ the equations become 
\begin{equation}
-\frac{\partial v_{i}^{\overrightarrow{(\mathcal{E})}}}{\partial t}-\frac{\partial v_{i}^{\overrightarrow{(\mathcal{E})}}}{\partial x^{a}}\left(v^{a}\right)^{\overrightarrow{(\mathcal{E})}}-\frac{\partial}{\partial x^{i}}\left(Q(p)\right)+\frac{1}{\rho}\frac{\partial}{\partial x^{b}}\mathsf{(\mathbb{T}^{\mathrm{cap}})}_{i}^{b}-\left(\frac{\partial U/\rho_{0}}{\partial\chi^{i}}\right)^{\overrightarrow{(\mathcal{E})}}=0.\label{Euler Gradient form}
\end{equation}

\paragraph{The calculation of $\frac{\partial}{\partial x^{a}}\mathsf{(\mathbb{T}^{cap})}_{i}^{a}$}

We have to compute the following term 
\begin{align*}
\frac{\partial}{\partial x^{a}}\mathsf{(\mathbb{T}^{\mathrm{cap}})}_{i}^{a} & =\frac{\partial}{\partial x^{a}}\left(\left(-\mathcal{P}^{\mathrm{cap}}\left(\rho,\beta\right)+\rho\frac{\partial}{\partial x^{b}}\left(2\frac{\partial\check{W}^{\mathrm{cap}}}{\partial\beta}g^{bc}\rho_{,c}\right)\right)\delta_{i}^{a}-2\frac{\partial\check{W}^{\mathrm{cap}}}{\partial\beta}g^{da}\rho_{,d}\rho_{,i}\right)\\
 & =\underset{A}{\underbrace{\frac{\partial}{\partial x^{i}}\left(\check{W}^{\mathrm{cap}}-\rho\frac{\partial\check{W}^{\mathrm{cap}}}{\partial\rho}+\rho\frac{\partial}{\partial x^{b}}\left(2\frac{\partial\check{W}^{\mathrm{cap}}}{\partial\beta}g^{bc}\rho_{,c}\right)\right)}}\underset{B}{\underbrace{-2\frac{\partial}{\partial x^{a}}\left(\frac{\partial\check{W}^{\mathrm{cap}}}{\partial\beta}g^{da}\rho_{,d}\rho_{,i}\right).}}
\end{align*}
Let us process first the term labeled $A$: 
\begin{gather*}
\begin{aligned}A & =\frac{\partial}{\partial x^{i}}\check{W}^{\mathrm{cap}}-\frac{\partial}{\partial x^{i}}\left(\rho\frac{\partial\check{W}^{\mathrm{cap}}}{\partial\rho}\right)+\frac{\partial}{\partial x^{i}}\left(\rho\frac{\partial}{\partial x^{b}}\left(2\frac{\partial\check{W}^{\mathrm{cap}}}{\partial\beta}g^{bc}\rho_{,c}\right)\right)\\
 & =\frac{\partial\check{W}^{\mathrm{cap}}}{\partial\rho}\frac{\partial\rho}{\partial x^{i}}+\frac{\partial\check{W}^{\mathrm{cap}}}{\partial\beta}\frac{\partial\beta}{\partial x^{i}}-\rho_{,i}\frac{\partial\check{W}^{\mathrm{cap}}}{\partial\rho}-\rho\frac{\partial}{\partial x^{i}}\left(\frac{\partial\check{W}^{\mathrm{cap}}}{\partial\rho}\right)+\frac{\partial}{\partial x^{i}}\left(\rho\frac{\partial}{\partial x^{b}}\left(2\frac{\partial\check{W}^{\mathrm{cap}}}{\partial\beta}g^{bc}\rho_{,c}\right)\right)\\
 & =\frac{\partial\check{W}^{\mathrm{cap}}}{\partial\beta}\frac{\partial\beta}{\partial x^{i}}-\rho\frac{\partial}{\partial x^{i}}\left(\frac{\partial\check{W}^{\mathrm{cap}}}{\partial\rho}\right)+\rho_{,i}\frac{\partial}{\partial x^{b}}\left(2\frac{\partial\check{W}^{\mathrm{cap}}}{\partial\beta}g^{bc}\rho_{,c}\right)+\left(\rho\frac{\partial}{\partial x^{i}}\frac{\partial}{\partial x^{b}}\left(2\frac{\partial\check{W}^{\mathrm{cap}}}{\partial\beta}g^{bc}\rho_{,c}\right)\right).
\end{aligned}
\end{gather*}
The term $B$ is easy to determine:

\[
B=-2\rho_{,i}\frac{\partial}{\partial x^{a}}\left(\frac{\partial\check{W}^{\mathrm{cap}}}{\partial\beta}g^{ad}\rho_{,d}\right)-2\frac{\partial\check{W}^{\mathrm{cap}}}{\partial\beta}g^{da}\rho_{,d}\rho_{,ia}
\]
Therefore we have 
\[
\begin{aligned}\frac{\partial}{\partial x^{a}}\mathsf{(\mathbb{T}^{\mathrm{cap}})}_{i}^{a} & =-\rho\frac{\partial}{\partial x^{i}}\left(\frac{\partial\check{W}^{\mathrm{cap}}}{\partial\rho}\right)+\left(\rho\frac{\partial}{\partial x^{i}}\frac{\partial}{\partial x^{b}}\left(2\frac{\partial\check{W}^{\mathrm{cap}}}{\partial\beta}g^{bc}\rho_{,c}\right)\right)+\frac{\partial\check{W}^{\mathrm{cap}}}{\partial\beta}\frac{\partial\beta}{\partial x^{i}}-2\frac{\partial\check{W}^{\mathrm{cap}}}{\partial\beta}g^{da}\rho_{,d}\rho_{,ia}\end{aligned}
,
\]
and recalling that 
\[
\frac{\partial\beta}{\partial x^{i}}=\frac{\partial}{\partial x^{i}}\left(g^{ab}\rho_{,a}\rho_{,b}\right)=2g^{ab}\rho_{,a}\rho_{,bi},
\]
the desired result is finally obtained 
\begin{align*}
\frac{\partial}{\partial x^{a}}\mathsf{(\mathbb{T}^{\mathrm{cap}})}_{i}^{a} & =-\rho\frac{\partial}{\partial x^{i}}\left(\frac{\partial\check{W}^{\mathrm{cap}}}{\partial\rho}\right)+\rho\frac{\partial}{\partial x^{i}}\frac{\partial}{\partial x^{b}}\left(2\frac{\partial\check{W}^{\mathrm{cap}}}{\partial\beta}g^{bc}\rho_{,c}\right)\\
 & =\rho\frac{\partial}{\partial x^{i}}\left(\frac{\partial}{\partial x^{b}}\left(2\frac{\partial\check{W}^{\mathrm{cap}}}{\partial\beta}g^{bc}\rho_{,c}\right)-\left(\frac{\partial\check{W}^{\mathrm{cap}}}{\partial\rho}\right)\right)\\
 & =\rho\frac{\partial}{\partial x^{i}}\left(\mathcal{P}^{\mathrm{eff}}\left(\rho;\rho_{,a};g^{ab}\rho_{,ab}\right)\right).
\end{align*}

\subsubsection{Bernoulli constant of motion along flow curves}

To conclude our argument we need a last tensorial equality (see e.g.
Lebedev et al. \cite{Lebedev}) 
\begin{equation}
\frac{\partial v_{i}}{\partial x^{a}}v^{a}=\frac{\partial v^{a}}{\partial x^{i}}v_{a}+\left(\frac{\partial v_{i}}{\partial x^{a}}v^{a}-\frac{\partial v^{a}}{\partial x^{i}}v_{a}\right)=\frac{\partial}{\partial x^{i}}\left(\frac{1}{2}v^{a}v_{a}\right)+W_{i}^{a}v_{a},\label{Stokes}
\end{equation}
where the tensor $W_{i}^{j}$ defined by 
\[
W_{i}^{j}:=\frac{\partial v_{i}}{\partial x^{j}}-\frac{\partial v^{j}}{\partial x^{i}}
\]
clearly saisfies the equality 
\[
W_{b}^{a}v_{a}v^{b}=\left(\frac{\partial v_{b}}{\partial x^{a}}v^{b}v^{a}-\frac{\partial v^{a}}{\partial x^{b}}v^{b}v_{a}\right)=\frac{1}{2}\left(\frac{\partial\left(v_{b}v^{b}\right)}{\partial x^{a}}v^{a}-\frac{\partial\left(v^{a}v_{a}\right)}{\partial x^{b}}v^{b}\right)=0.
\]
Let consider the equations \eqref{Euler Gradient form} 
\[
-\frac{\partial v_{i}^{\overrightarrow{(\mathcal{E})}}}{\partial t}-\frac{\partial v_{i}^{\overrightarrow{(\mathcal{E})}}}{\partial x^{a}}\left(v^{a}\right)^{\overrightarrow{(\mathcal{E})}}-\frac{\partial}{\partial x^{i}}\left(Q(p)\right)+\frac{1}{\rho}\frac{\partial}{\partial x^{b}}\left(\mathsf{\mathbb{S}}_{i}^{b}\right)-\left(\frac{\partial U/\rho_{0}}{\partial\chi^{i}}\right)^{\overrightarrow{(\mathcal{E})}}=0.
\]
If the applied bulk external forces are such that there exists a scalar
Eulerian function $V$ for which 
\[
\left(\frac{\partial U/\rho_{0}}{\partial\chi^{i}}\right)^{\overrightarrow{(\mathcal{E})}}=\frac{\partial V}{\partial x^{i}},
\]
and by making use of \eqref{Stokes} ( the notation $\left(\cdot\right)^{\overrightarrow{(\mathcal{E})}}$
has been dropped), we obtain 
\[
-\frac{\partial v_{i}}{\partial t}-\frac{\partial}{\partial x^{i}}\left(\frac{1}{2}v^{c}v_{c}\right)+W_{i}^{d}v_{d}-\frac{\partial}{\partial x^{i}}\left(Q(p)\right)+\frac{\partial}{\partial x^{i}}\left(\frac{\partial}{\partial x^{b}}\left(2\frac{\partial\check{W}^{\mathrm{cap}}}{\partial\beta}g^{ab}\rho_{,a}\right)-\left(\frac{\partial\check{W}^{\mathrm{cap}}}{\partial\rho}\right)\right)-\frac{\partial V}{\partial x^{i}}=0.
\]
By calculating the inner product with $v$ we get 
\[
\frac{\partial}{\partial t}\left(\frac{1}{2}v\cdot v\right)+\nabla\left(\frac{1}{2}v\cdot v+Q(p(\rho))-\mathcal{P}^{\mathrm{eff}}\left(\rho;\rho_{,a};g^{ab}\rho_{,ab}\right)+V\right)\cdot v=0,
\]
and if the field $v$ be stationary, i.e. if 
\[
\frac{\partial v}{\partial t}=0,
\]
the last equation becomes 
\[
\nabla\left(\frac{1}{2}v\cdot v+Q(p(\rho))-\mathcal{P}^{\mathrm{eff}}\left(\rho;\rho_{,a};g^{ab}\rho_{,ab}\right)+V\right)\cdot v=0
\]
i.e. along (steady) flow curves there exists a constant $K_{0}$ such
that 
\[
\frac{1}{2}v\cdot v+Q(p(\rho))-\mathcal{P}^{\mathrm{eff}}\left(\rho;\rho_{,a};g^{ab}\rho_{,ab}\right)+V=K_{0}.
\]

\section{Conclusions:\ towards continuum analytical mechanics ?}

The role of the principle of least action (or of its weaker version
the principle of virtual work) in applied mathematics, and in particular
in mathematical physics, has been controversial since its very first
formulations. The attitude towards this postulation is often one of
total rejection. Indeed, both the supporters of \emph{variational}
postulations and the supporters of \emph{balance of everything} behave
often as if the controversy does not exist. They simply pretend that
the other postulation process is not used at all or anymore. Of course
the supporters of \emph{balance of everything} are aware of the importance
of a variational principle, especially when a numerical code has to
be designed or an existence and uniqueness theorem needs to be proved.
They treat the variational principle as a theorem to be proved in
their postulation scheme. Very strange and somehow clumsy expressions
are used like:\ \emph{theorem of the principle of virtual work} which
is rather an oxymoron. Their attitude (see the section on \emph{variational
principles} in Truesdell and Toupin \cite{TruesdellToupin}) is that
a variational formulation cannot be generally obtained. If they exist,
they are considered as mathematical curiosities that merely facilitate
the work of the mathematicians. For them the search for variational
principles is a secondary task relegated to the applied mathematicians.

On the contrary the supporters of \emph{variational} postulations
behave as if their point of view were the only one possible: they
do not even care to announce that they use it as, in their opinion,
everybody has to do so. To these supporters are directed the words
of Piola which we already cited:

\textit{``Somebody could here object that this }{[}i.e. the variational
foundations of Analytical Mechanics{]} \textit{is a very old knowledge,
which does not deserve to be newly promulgated by me: however {[}it
seems that my efforts are needed{]} as my beautiful theories {[}after
being published{]} are then criticized.''}

Actually the elitist attitude of many supporters of \emph{variational}
postulations is the true cause of the frequent \emph{rediscoveries}
of the same variational principles in different times and the loss
of the information about their first historical appearance. Variational
principles have to be regarded as the most powerful heuristic tool
in applied mathematics. The wise attitude of Hamilton and Rayleigh
consisted in refraining from the effort of describing dissipative
phenomena directly and explicitly by means of the least action principle,
but including them in the picture only in a second step, by means
of the introduction of a suitable dissipation functional. Of course
this \emph{heuristic} attitude does not imply that a purely variational
formulation of given model cannot be obtained, at worst by embedding
the original space of configurations in a wider one. When this further
step can be performed then the value of the improved mathematical
model will increase.

In this context we found interesting the works Carcaterra and Sestieri
\cite{CarcaterraSestieri1995}, Carcaterra et al. \cite{CarcaterraCiappi2000},
Culla et al. \cite{CullaSestieriCarcaterra2003}, Carcaterra \cite{Carcaterra2005},
Carcaterra and Akai \cite{CarcaterraAkay2007}, which were initially
motivated by the need to develop innovative technological solutions.
In these papers it is proven that a conservative system can show,
if one restricts his attention to a subset of its degrees of freedom,
an apparent dissipative behavior. Actually in suitably designed conservative
systems the energy may flow from some primary degrees of freedom into
a precise set of other (secondary or hidden) ones, and remain there
trapped for a very long (from the point of view of practical application:
infinite) time. Therefore, in some cases, a non-conservative description
of a primary system, including an \textit{ad-hoc} dissipation functional,
is a realistic and effective modeling simplification, even if the
\emph{true} and complete system is actually Hamiltonian and conservative.
The greatest advantage in variational based models is that, if the
action functional is well-behaved, they always produce intrinsically
well-posed mathematical problems. Somebody claimed that this is a
purely mathematical requirement: actually this is not the case. \emph{It
is a \textquotedbl{}physical\textquotedbl{} prescription that a model
could give a \textquotedbl{}unique\textquotedbl{} provision of the
modalities of occurrence of a physical phenomenon!}

There is also a \emph{practical} advantage in the variational formulation
of models as they are easily transformed into numerical codes. Of
course after having considered Lagrangian systems (the evolution of
which are governed by a least action functional) the study of non-Lagrangian
ones (for which such a functional may not exist) may appear very difficult.
It is often stated that dissipation cannot be described by means of
a least action principle. This is not exactly true, as it is possible
to find some action functionals for a large class of dissipative systems
(see e.g. Maugin \cite{Maugin2000},Vujanovic and Jones \cite{VujanovicJones1989}
or Moiseiwitsch \cite{Moisewitsch}). However it is true that not
every conceived system can be regarded as a Lagrangian one. This point
is mathematically delicate and will be only evoked here (see e.g.
Santilli \cite{Santilli} for more details). In general, a non-Lagragian
system can be regarded as Lagrangian in two different ways: i) because
it is an \emph{approximation} of a Lagrangian system (see the case
of Cattaneo equation for heat propagation in e.g. Vujanovic and Jones\cite{VujanovicJones1989}),
and this approximation leads to \emph{cancel} the lacking part of
the \emph{true} action functional; ii) because the considered system
is simply a subsystem of a larger one which is truly Lagrangian. (see
e.g. Carcaterra and Sestieri \cite{CarcaterraSestieri1995}, Carcaterra
et al. \cite{CarcaterraCiappi2000} Carcaterra \cite{Carcaterra2005},
Carcaterra ans Akai\cite{CarcaterraAkay2007} \cite{CarcaterraAkay2007}).
The assumption that variational principles can be used only for non-dissipative
systems is contradicted by, e.g., Bourdin et al.\cite{BurdinFrancfortMarigo2008},
Maugin and Trimarco \cite{MauginTrimarco1992} or Rinaldi and Lai
\cite{RinaldiLai2007} where variational principles modeling dissipative
phenomena occurring in damage and fracture are formulated. In our
opinion models for surface phenomena in presence of thermodynamical
phenomena and diffusion or phase transitions in solids developed e.g.
in McBride et al \cite{Mac,Mac1}, Steeb and Diebels \cite{DiebelsStefan}
and Steinmann et al. \cite{Steinmann} or for growth phenomena in
living tissues such as those presented in \cite{MadeoLekszyckidell'isola2011}
(with suitable modifications!) should be formulated in a variational
form.

One should not believe that the aforementioned considerations are
limited to the description of mechanical phenomena only: actually
the formulation of \emph{variational principles} proved to be a powerful
tool in many different research fields. In the following list (which
cannot be exhaustive) we simply want to indicate the enormous variety
of phenomena which were considered, up to now, from the variational
point of view, by citing only those few works among the many available
in the literature that are more familiar to us:
\begin{itemize}
\item for biological evolutionary phenomena (see e.g. Edwards \cite{Edwards},
Klimek et al. \cite{KlimekThurnerHanel2010} and references therein); 
\item for the mathematical study of mutation and selection phenomena in
species evolution (see e.g. Baake and Georgii \cite{BaakeGeorgii}); 
\item for some phenomena of solid/solid phase transitions in plates and
shells (see e..g. to Eremeyev Pietraszkiewicz et al. \cite{Pietraszkiewicz WEremeyevKonopinska2007},
Eremeev et al. \cite{EremeevFreidinSharipova2003}, Eremeyev and Pietraszkiewicz
\cite{Eremeyev Pietraszkiewicz2004} ); 
\item for mechanical vibration control (see e.g. Carcaterra and Akai \cite{CarcaterraAkay2007}); 
\item for electromagnetic phenomena (see e.g. Daher and Maugin \cite{DaherMaugin1986b}
and references therein); 
\item for vibration control using distributed arrays of piezoelectric actuators
(see e.g. dell'Isola Vidoli \cite{dellisolavidoli1998,dellisolavidoli1998AAM}); 
\item for interfacial phenomena (see e.g. Eremeyev and Pietraszkiewicz \cite{Eremeyev Pietraszkiewicz2004},
\cite{Cuomo1} Steigmann and Ogden \cite{Steigmann6}, Daher and Maugin
\cite{DaherMaugin1986} and references therein); 
\item for the theory of membranes and rods (see e.g. Steigmann and Faulkner
\cite{Steigmann9}); 
\item for mechanical phenomena involving different length scales (see e.g.
Steigmann \cite{Steigmann8}, dell'Isola et al. \cite{dellisolamadeoseppecher2009}
and references therein); 
\item for phase transition phenomena in fluids (see Seppecher \cite{GatignolSeppecher,SEPPECHER1987,SEPPECHER1988,SEPPECHER1989,SEPPECHER1993}
or Casal and Gouin \cite{Casal-Gouin1985,CASALGOUIN1988}); 
\item for damage and fracture phenomena (see e.g. Francfort and Marigo \cite{FrancfortMarigo1998},
Yang and Misra \cite{MISRA2,MISRA3}, Contrafatto and Cuomo \cite{Contrafatto1,Contrafatto2,Contrafatto3},
\cite{Cuomo2}, Rinaldi and Lai \cite{RinaldiLai2007} and Del Piero
\cite{delpiero2013}) ; 
\item for some phenomena related to fluid flow in deformable porous media
(see e.g. to dell'Isola et al. \cite{dellisolaguarasciohutter(2000)},
dell'Isola et al. \cite{dellisolamadeoseppecher2009}, Sciarra et
al. \cite{SciarradellisolaHutter,SciarradellisolaCoussy,SciarradellisolaIaniroMadeo},
Quiligotti et al. \cite{QUILIGOTTIMAUGINdellisola(2003)}); 
\item for some piezoelectromechanical or magnetoelastic coupling phenomena
(see e.g. to Barham et al. \cite{Barham}, Maurini et al. \cite{Maurinidellisoladelvescovo2004},
Maugin and Attou \cite{MauginAttou1989}, Maurini, et al. \cite{MauriniPougetdellisola2004},
dell'Isola and Vidoli \cite{dellisolavidoli1998,dellisolavidoli1998AAM}). 
\end{itemize}
\bigskip{}

\section{Acknowledgments}

The senior author F.d.I. would like to thank his students of the course
``Meccanica Analitica per Fisici'' which he taught in the Academic
Year 2010/2011 at the Università di Napoli ``Federico II'' (his
Alma Mater) and the students of the Doctoral School in Theoretical
and Applied Mechanics of the Università di Roma ``La Sapienza''.
Their demanding attitude towards the Professor obliged him -after
having searched unsuccessfully in the literature- to write a paper
where he had to prove that the Lagragian principle of least action
can be the basis of the study of capillary fluids also. The ideas
expressed by Pierre Seppecher during years of collaboration also greatly
influenced this paper, even if it is not sure that he will approve
all presented conclusions. Also the fruitful discussions with Prof.
Carlo Massimo Casciola were very helpful.

This work was supported by the International Research Center M\&MoCS.
V.A.E. was supported by the RFBR {[}grant number 12-01-00038{]}. A.M.
was supported by the project BQR 2013 \textquotedblleft{}Matériaux
Méso et Micro-Hétérogènes: Optimisation par Modèles de Second Gradient
et Applications en Ingénierie\textquotedblright{} {[}BQR 2013-0054{]}.

\appendix
%dummy comment inserted by tex2lyx to ensure that this paragraph is not empty
%dummy comment inserted by tex2lyx to ensure that this paragraph is not empty
%dummy comment inserted by tex2lyx to ensure that this paragraph is not empty
%dummy comment inserted by tex2lyx to ensure that this paragraph is not empty
%dummy comment inserted by tex2lyx to ensure that this paragraph is not empty
%dummy comment inserted by tex2lyx to ensure that this paragraph is not empty

\section{Piola transformations and the formula of material derivative}

\subsection{Geometric framework}

Let $\chi$ be a $C^{2}$-diffeomorphism between the domains $D_{\alpha}$
and $D_{\beta}.$ The following notations will be adopted 
\[
F:=\nabla\chi,\qquad J:=\det F,\qquad F^{-T}:=\left(F^{-1}\right)^{T}
\]
These fields are all defined in $D_{\alpha}$. Conversely, the fields
\[
F^{-1},\qquad J^{-1}:=\det F^{-1},\qquad F^{T}
\]
are obviously defined on $D_{\beta}.$ These relations are summed
up in the following diagram: 
\[
\xyC{5pc}\xyR{2pc}\xymatrix{T_{X}D_{\alpha}\ar@<2pt>[r]^{F} & T_{x}D_{\beta}\ar@<2pt>[l]^{F^{T}}\\
D_{\alpha}\ar[r]^{\chi}\ar[u]\ar[d] & D_{\beta}\ar[u]\ar[d]\\
T_{X}^{\star}D_{\alpha}\ar@<2pt>[r]^{F^{-T}} & T_{x}^{\star}D_{\beta}\ar@<2pt>[l]^{F^{-1}}
}
\]
in which $T_{p}D$ and $T_{p}^{\star}D$ denote, respectively, the
tangent and cotangent plane to $D$ at $p$. For every tensor field
$T_{\alpha}$ defined in $D_{\alpha}$, and for every tensor field
$T_{\beta}$ defined in $D_{\beta}$, we use the notations 
\[
T_{\alpha}^{\overrightarrow{(\beta)}}:=T_{\alpha}\circ\chi^{-1},\qquad T_{\beta}^{\overrightarrow{(\alpha)}}:=T_{\beta}\circ\chi.
\]
We will say that $T_{\alpha}^{\overrightarrow{(\beta)}}$ is the field
$T_{\alpha}$ displaced in $D_{\beta}$ and conversely. These relations
are exemplified in the following diagram in the specific case of two
vectors fields:

\[
\xyC{5pc}\xyR{6pc}\xymatrix{T_{X}D_{\alpha}\ar@<2pt>[r]^{F} & T_{x}D_{\beta}\ar@<2pt>[l]^{F^{T}}\\
D_{\alpha}\ar@<2pt>[r]^{\chi}\ar[u]^{T_{\alpha}}\ar[ru]\sp(0.3){T_{\beta}^{\overrightarrow{(\alpha)}}} & D_{\beta}\ar[u]^{T_{\beta}}\ar@<2pt>[l]^{\chi^{-1}}\ar[lu]\sb(0.3){T_{\alpha}^{\overrightarrow{(\beta)}}}
}
\]

\subsection{Transposition of linear mappings}

The transposed $F^{T}$ of the linear mapping $F$ from the vector
space $T_{X}D_{\alpha}$ to the vector space $T_{x}D_{\beta}$ is
defined as the unique linear mapping from $T_{x}^{\star}D_{\beta}$
to $T_{X}^{\star}D_{\alpha}$ such that for every couple $\left(V,l\right)\in T_{X}D_{\alpha}\times T_{x}^{\star}D_{\beta}$
\[
\left\langle l,FV\right\rangle _{(T_{x}^{\star}D_{\beta},T_{x}D_{\beta})}=\left\langle F^{T}l,V\right\rangle _{(T_{X}^{\star}D_{\alpha},T_{X}D_{\alpha})}
\]
where the bracket denotes the duality product. If both $D_{\alpha}$
and $D_{\beta}$ are equipped with an inner product on their tangent
space at each point%
\footnote{In others terms, if both $D_{\alpha}$ and $D_{\beta}$ are Riemannian
manifolds.%
}, then tangent and cotangent space can be identified. Let us denote
by $g_{\alpha}$ and $g_{\beta}$ the fields of metric defined, respectively,
on $D_{\alpha}$ and $D_{\beta}$. Through $g_{\beta}$ a vector $w$
can be associated to any covector $l$, more precisely:

\[
\forall l\epsilon T_{x}^{\star}D_{\beta},\,\exists\, w\,\epsilon\, T_{x}D_{\beta},\quad l=g_{\beta}w
\]
Therefore the equality between the duality bracket can be rewritten

\[
\left\langle g_{\beta}w,FV\right\rangle _{(T_{x}^{\star}D_{\beta},T_{x}D_{\beta})}=\left\langle F^{T}g_{\beta}w,V\right\rangle _{(T_{X}^{\star}D_{\alpha},T_{X}D_{\alpha})}
\]
This construction can be summarized by the following diagram 
\[
\xyC{3pc}\xyR{3pc}\xymatrix{ & T_{X}^{\star}D_{\alpha}\ar[ld] & T_{x}^{\star}D_{\beta}\ar[l]^{F^{T}}\ar[rd]\\
\mathbb{R} &  &  & \mathbb{R}\\
\mathbb{} & T_{X}D_{\alpha}\ar[r]^{F}\ar[uu]^{g_{\alpha}}\ar[lu] & T_{x}D_{\beta}\ar[uu]^{g_{\beta}}\ar[ru] & \mathbb{}
}
\]
Once bases are introduced in $T_{X}D_{\alpha}$ and $T_{x}D_{\beta}$,
we can represent vectors, tensors and inner products in terms of their
components. The following relation is written in the domain $D_{\beta}$,
hence quantities defined on $D_{\alpha}$ have to be transported:
\[
g_{ab}w^{b}(F_{A}^{a}V^{A})\overrightarrow{^{(\beta)}}=V^{A}\left(F^{T}\right)_{A}^{a}g_{ab}w^{b}(V^{A})\overrightarrow{^{(\beta)}}\qquad\forall V^{A},\forall w^{b},
\]
which implies 
\[
g_{ab}\left((F_{A}^{a})\overrightarrow{^{(\beta)}}-\left(F^{T}\right)_{A}^{a}\right)V^{A}w^{b}=0\qquad\forall V^{A},\forall w^{b}.
\]
Therefore we have 
\begin{equation}
(F_{M}^{i})\overrightarrow{^{(\beta)}}=\left(F^{T}\right)_{M}^{i},\label{eq:DepGra}
\end{equation}
and, conversely,

\begin{equation}
(F^{-1})_{i}^{M}\overrightarrow{^{(\alpha)}}=\left(F^{-T}\right)_{i}^{M}.\label{eq:DepInvGra}
\end{equation}
These relations will be important in the next subsection to properly
define the Piola transformation.

Let us now consider the following inner product (with a slight abuse
of notation) 
\[
\left\langle FV,FW\right\rangle _{T_{x}D_{\beta}},
\]
where $F$ is the same linear mapping as before. By considering the
transposed mapping one gets 
\[
\left\langle FV,FW\right\rangle _{T_{x}D_{\beta}}=\left\langle F^{T}FV,W\right\rangle _{T_{X}D_{\alpha}},
\]
which in terms of components becomes 
\[
(g_{ab})\overrightarrow{^{(\alpha)}}F_{A}^{a}V^{A}F_{B}^{b}W^{B}=g_{CB}(F^{T}F)_{B}^{C}V^{A}W^{B},
\]
therefore 
\[
(F^{T}F)_{MN}=(g_{ab})\overrightarrow{^{(\alpha)}}F_{M}^{a}F_{N}^{b}=(F_{Ma})\overrightarrow{^{(\alpha)}}F_{N}^{a},
\]
or more simply, dropping the change of domain:

\[
(F^{T}F)_{MN}=F_{Ma}F_{N}^{a}\quad;\quad(F^{T}F)^{MN}=F_{a}^{M}F^{aN}.
\]

\subsection{Piola transformation for virtual work and stress tensors}

We call \emph{virtual displacement stemming from} $\chi$ a vector
field $\delta\chi$ defined in $D_{\alpha}$ and such that, for every
$X$ in $D_{\alpha}$, the vector $\delta\chi(X)$ belongs to the
tangent space at the point $\chi(X).$ We will denote by $\mathcal{D}$
the space of such virtual displacements: 
\[
\mathcal{D}=\{\delta\chi:D_{\alpha}\rightarrow TD_{\beta},X\mapsto\delta\chi(X)\}.
\]
A virtual work functional must obviously be identified as a linear
and continuous functional defined on $\mathcal{D}$ (for a detailed
discussion of this point see dell'Isola et al. \cite{dellisolaseppechermadeo2012}
and references therein), i.e to an element of $\mathcal{D}^{\star}$
the dual space of $\mathcal{D}$ 
\[
\mathcal{D}^{\star}=\{\mathcal{W}:\mathcal{D}\rightarrow\mathbb{R},\delta\chi\mapsto W\}.
\]
Because of a representation theorem due to Schwartz \cite{Sch} we
can state that for any virtual work functional $\mathcal{W}$ defined
in $D_{\alpha}$ there exist $N$ regular fields $\underset{\gamma}{\mathrm{P}}$
(where $\gamma=1,...N$) such that 
\[
\mathcal{W}\left(\delta\chi\right)={\displaystyle \sum\limits _{\gamma=1}^{N}}{\displaystyle \int\limits _{D_{\alpha}}}\underset{{\scriptstyle \gamma}}{\mathrm{P}}\underset{{\scriptstyle \gamma}}{\nabla_{\alpha}}\left(\mathcal{\delta\chi}\right)dV_{\alpha},
\]
where 
\[
\underset{{\scriptstyle \gamma}}{\nabla_{\alpha}}=D_{\alpha}\rightarrow\otimes^{\gamma}T^{\star}D_{\alpha}\otimes TD_{\beta}\quad;\quad\underset{{\scriptstyle \gamma}}{\mathrm{P}}=D_{\alpha}\rightarrow\otimes^{\gamma}TD_{\alpha}\otimes T^{\star}D_{\beta}.
\]
Modifying slightly the nomenclature introduced by Truesdell and Toupin
\cite{TruesdellToupin} we can call $\underset{\gamma}{\mathrm{P}}$
the $\gamma-th$ order Piola stress tensor. Now, following Piola \cite{Piola4},
we can transport the field $\mathcal{\delta\chi}$ on $D_{\beta}$
and define the corresponding Cauchy stress tensors $\underset{{\scriptstyle \gamma}}{\mathrm{T}}$
by means of the equality 
\[
{\displaystyle \int\limits _{D_{\alpha}}}\underset{{\scriptstyle \gamma}}{\mathrm{P}}\underset{{\scriptstyle \gamma}}{\nabla_{\alpha}}\left(\mathcal{\delta\chi}\right)dV_{\alpha}:=\int\limits _{D_{\beta}}\underset{{\scriptstyle \gamma}}{\mathrm{T}}\underset{{\scriptstyle \gamma}}{\nabla_{\beta}}\left(\mathcal{\delta\chi}^{\overrightarrow{(\beta)}}\right)dV_{\beta}\qquad\forall\mathcal{\delta\chi}\in\mathcal{D},
\]
in which 
\[
\underset{{\scriptstyle \gamma}}{\nabla_{\beta}}=D_{\beta}\rightarrow\otimes^{\gamma}T^{\star}D_{\beta}\otimes TD_{\beta}\quad;\quad\underset{{\scriptstyle \gamma}}{\mathrm{T}}=D_{\alpha}\rightarrow\otimes^{\gamma}TD_{\beta}\otimes T^{\star}D_{\beta}.
\]
To prove that such a tensor exists, and to get its representation,
let us write component-wise the previous equation: 
\[
{\displaystyle \int\limits _{D_{\alpha}}}\mathrm{P}_{i}^{A_{1}...A_{\gamma}}\left(\mathcal{\delta\chi}\right)_{,A_{1}...A_{\gamma}}^{i}dV_{\alpha}={\displaystyle \int\limits _{D_{\beta}}}\mathrm{T}_{i}^{j_{1}....j_{\gamma}}\left(\mathcal{\delta\chi}^{\overrightarrow{(\beta)}}\right)_{,j_{1}....j_{\gamma}}^{i}dV_{\beta}\qquad\forall\mathcal{\delta\chi}\in\mathcal{D}.
\]
Then using the chain rule the derivatives: 
\[
\left(\mathcal{\delta\chi}^{\overrightarrow{(\beta)}}\right)_{,j_{1}....j_{\gamma}}^{i}=\left(\left(\mathcal{\delta\chi}\right)_{,A_{1}...A_{\gamma}}^{i}\right)^{\overrightarrow{(\beta)}}\left(F^{-1}\right)_{j_{1}}^{A_{1}}...\left(F^{-1}\right)_{j_{\gamma}}^{A_{\gamma}},
\]
and a change of variable in the second integral, we obtain 
\[
{\displaystyle \int\limits _{D_{\alpha}}}\mathrm{P}_{i}^{A_{1}...A_{\gamma}}\left(\mathcal{\delta\chi}\right)_{,A_{1}...A_{\gamma}}^{i}dV_{\alpha}={\displaystyle \int\limits _{D_{\alpha}}}J\left(\mathrm{T}_{i}^{j_{1}....j_{\gamma}}\left(F^{-1}\right)_{j_{1}}^{A_{1}}...\left(F^{-1}\right)_{j_{\gamma}}^{A_{\gamma}}\right)^{\overrightarrow{(\alpha)}}\left(\mathcal{\delta\chi}\right)_{,A_{1}...A_{\gamma}}^{i}dV_{\alpha}\qquad\forall\mathcal{\delta\chi}\in\mathcal{D},
\]
which is equivalent to the following Piola formula for transformation
of stress tensors

\[
\mathrm{P}_{i}^{A_{1}...A_{\gamma}}=J\left(\mathrm{T}_{i}^{j_{1}....j_{\gamma}}\left(F^{-1}\right)_{j_{1}}^{A_{1}}...\left(F^{-1}\right)_{j_{\gamma}}^{A_{\gamma}}\right)^{\overrightarrow{(\alpha)}};
\]
or, using the transformation \eqref{eq:DepInvGra} :

\[
\mathrm{\underset{{\scriptstyle \gamma}}{\mathrm{P}}}=J\left(\underset{{\scriptstyle \gamma}}{\mathrm{T}}\right)^{\overrightarrow{(\alpha)}}\underset{\gamma}{\underbrace{F^{-T}\ldots F^{-T}}}.
\]
With simple algebra we also get 
\[
J^{-1}\left(\mathrm{P}_{i}^{A_{1}...A_{\gamma}}F_{A_{1}}^{i_{1}}....F_{A_{\gamma}}^{i_{\gamma}}\right)^{\overrightarrow{(\beta)}}=\mathrm{T}_{\gamma i}^{i_{1}....i_{\gamma}}
\]
or, using the transformation \eqref{eq:DepGra} :

\[
\mathrm{\underset{{\scriptstyle \gamma}}{T}}=J^{-1}\left(\underset{{\textstyle {\scriptstyle \gamma}}}{P}\right)^{\overrightarrow{(\beta)}}\underset{\gamma}{\underbrace{F^{T}\ldots F^{T}}}.
\]

\subsection{Piola transformation for divergence}

For any tensor field $T_{\alpha}$ the following equality holds (for
a proof see e.g. dell'Isola et al. \cite{dellisolaplacidi2011} or
Hughes and Marsden \cite{MarHu}). 
\begin{equation}
\nabla_{\alpha}\cdot T_{\alpha}=J\ \left(\nabla_{\beta}\cdot\left(J^{-1}T_{\alpha}^{\overrightarrow{(\beta)}}F^{T}\right)\right)^{\overrightarrow{(\alpha)}}\label{Transformation Formula}
\end{equation}
which obviously implies, vice versa, 
\[
\left(\nabla_{\alpha}\cdot T_{\alpha}\right)^{\overrightarrow{(\beta)}}=J^{\overrightarrow{(\beta)}}\ \nabla_{\beta}\cdot\left(J^{-1}T_{\alpha}^{\overrightarrow{(\beta)}}F^{T}\right).
\]
In components this relation reads (where $X^{L}$ and $x^{j}$ denote
the components of the position vector in $D_{\alpha}$ and $D_{\beta}$
respectively) 
\begin{equation}
\left(\frac{\partial T_{\alpha}^{A}}{\partial X^{A}}\right)^{\overrightarrow{(\beta)}}=J^{\overrightarrow{(\beta)}}\ \frac{\partial}{\partial x^{a}}\left(J^{-1}\left(T_{\alpha}^{A}F_{A}^{a}\right)^{\overrightarrow{(\beta)}}\right).\label{PIOLATRANSFORMATION}
\end{equation}
Similarly we have that the following relationship, in some sense inverse
of the relation \eqref{Transformation Formula}: 
\begin{equation}
\nabla_{\beta}\cdot T_{\beta}=J^{-1}\ \left(\nabla_{\alpha}\cdot\left(JT_{\beta}^{\overrightarrow{(\alpha)}}F^{-T}\right)\right)^{\overrightarrow{(\beta)}}.\label{DIVTBETA}
\end{equation}

\subsection{The Piola-Ricci-Bianchi condition}

The equation \eqref{DIVTBETA} was first found, \emph{without} the
help of tensor calculus, by Piola \cite{Piola4}. In the case where
$T_{\beta}$ reduces to the identity, the former equation takes the
following form 
\begin{equation}
\nabla\cdot\left(JF^{-T}\right)=0,\label{PiolaCondition}
\end{equation}
which in components can be written 
\[
\frac{\partial}{\partial X^{A}}\left(J\left(F^{-1}\right)_{i}^{A}\right)=0
\]
Equation \eqref{PiolaCondition} is a particular case of the Bianchi
condition for the Ricci curvature tensor, when interpreting Lagrangian
coordinates as a chart for the Eulerian configuration of the body.
Fromthe Piola-Ricci-Bianchi condition 
\[
\frac{\partial}{\partial X^{A}}\left(J\left(F^{-1}\right)_{i}^{A}\right)=0
\]
one gets

\[
\begin{aligned}J_{,A}\left(F^{-1}\right)_{i}^{A}+J\left(F^{-1}\right)_{i,A}^{A} & =0\\
\left(F^{-1}\right)_{i,A}^{A} & =-J^{-1}\left(\frac{\rho_{0}}{\rho}\right)_{,A}\left(F^{-1}\right)_{i}^{A}=-\rho_{0}J^{-1}\left(-\frac{1}{\rho^{2}}\right)\rho_{,A}\left(F^{-1}\right)_{i}^{A}=\left(\frac{1}{\rho}\right)\rho_{,A}\left(F^{-1}\right)_{i}^{A}.
\end{aligned}
\]
In conclusion 
\begin{equation}
\left(F^{-1}\right)_{i,A}^{A}=\frac{\rho_{,A}}{\rho}\left(F^{-1}\right)_{i}^{A}=\frac{\rho_{,i}}{\rho}.\label{CorollaryPiolaCondition}
\end{equation}

\subsection{Piola transformation for double divergence}

To obtain the Eulerian form for balance equation of the capillary
fluids we need to apply the divergence twice to calculate the transformation
of double Lagrangian divergence. We proceed as follows: the equality
\eqref{PIOLATRANSFORMATION} implies that (remark: we assume that
the tensor $T_{\alpha}^{AB}$ is symmetric) 
\[
\left(\frac{\partial T_{\alpha}^{AB}}{\partial X^{B}}\right)^{\overrightarrow{(\beta)}}=J^{\overrightarrow{(\beta)}}\ \frac{\partial}{\partial x^{b}}\left(J^{-1}\left(T_{\alpha}^{AB}F_{B}^{b}\right)^{\overrightarrow{(\beta)}}\right)
\]
Then 
\begin{align*}
\left(\frac{\partial}{\partial X^{A}}\left(\frac{\partial T_{\alpha}^{AB}}{\partial X^{B}}\right)\right)^{\overrightarrow{(\beta)}} & =J^{\overrightarrow{(\beta)}}\ \frac{\partial}{\partial x^{a}}\left(J^{-1}\left(\left(\frac{\partial T_{\alpha}^{AB}}{\partial X^{B}}\right)^{\overrightarrow{(\beta)}}\left(F_{A}^{a}\right)^{\overrightarrow{(\beta)}}\right)\right)\\
 & =J^{\overrightarrow{(\beta)}}\ \frac{\partial}{\partial x^{a}}\left(J^{-1}\left(J^{\overrightarrow{(\beta)}}\ \frac{\partial}{\partial x^{b}}\left(J^{-1}\left(T_{\alpha}^{AB}F_{B}^{b}\right)^{\overrightarrow{(\beta)}}\right)\left(F_{A}^{a}\right)^{\overrightarrow{(\beta)}}\right)\right)\\
 & =J^{\overrightarrow{(\beta)}}\ \frac{\partial}{\partial x^{a}}\ \left(\frac{\partial}{\partial x^{b}}\left(J^{-1}\left(T_{\alpha}^{AB}F_{B}^{b}\right)^{\overrightarrow{(\beta)}}\right)\left(F_{A}^{a}\right)^{\overrightarrow{(\beta)}}\right).
\end{align*}
In conclusion we have: 
\[
\left(\frac{\partial}{\partial X^{A}}\left(\frac{\partial T_{\alpha}^{AB}}{\partial X^{B}}\right)\right)^{\overrightarrow{(\beta)}}=J^{\overrightarrow{(\beta)}}\ \frac{\partial}{\partial x^{a}}\ \left(\frac{\partial}{\partial x^{b}}\left(J^{-1}\left(T_{\alpha}^{AB}F_{B}^{b}\right)^{\overrightarrow{(\beta)}}\right)\left(F_{A}^{a}\right)^{\overrightarrow{(\beta)}}\right).
\]

\subsection{Piola transformation for normals}

For normals we have the following formula (see e.g. dell'Isola et
al. \cite{dellisolamadeoseppecher2009}) 
\begin{equation}
N_{\alpha}^{\overrightarrow{(\beta)}}=\frac{\left(J^{-1}F^{T}\right)N_{\beta}}{\left\Vert \left(J^{-1}F^{T}\right)N_{\beta}\right\Vert }\label{TRANSFORMATIONOFNORMALS}
\end{equation}
while, for the passage from $\alpha$ to $\beta$ domain, the following
transformation formula for areas holds: 
\begin{equation}
\left(\left\Vert \left(J^{-1}F^{T}\right)N_{\beta}\right\Vert ^{-1}\right)^{\overrightarrow{(\alpha)}}=\left\Vert \left(JF^{-T}\right)N_{\alpha}\right\Vert =\frac{dA_{\beta}}{dA_{\alpha}}.\label{AREASTRANFORMATION}
\end{equation}

\subsection{Material derivative}

For the formula of material derivative we start by remarking that
\[
\left(T_{\alpha}^{\overrightarrow{(\beta)}}\right)^{\overrightarrow{(\alpha)}}=T_{\alpha}.
\]
Therefore 
\begin{align*}
\left(\left.\frac{\partial T_{\alpha}}{\partial t}\right\vert _{X}\right) & =\left(\left.\frac{\partial\left(T_{\alpha}^{\overrightarrow{(\beta)}}\right)^{\overrightarrow{(\alpha)}}}{\partial t}\right\vert _{X}\right)=\left(\left.\frac{\partial\left(T_{\alpha}^{\overrightarrow{(\beta)}}\circ\chi\right)}{\partial t}\right\vert _{X}\right)=\left(\left.\frac{\partial\left(T_{\alpha}^{\overrightarrow{(\beta)}}\left(\chi(X,t\right),t)\right)}{\partial t}\right\vert _{X}\right)\\
 & =\left(\left.\frac{\partial\left(T_{\alpha}^{\overrightarrow{(\beta)}}(x,t)\right)}{\partial t}\right\vert _{x}\circ\chi\right)+\left(\left.\nabla_{x}T_{\alpha}^{\overrightarrow{(\beta)}}(x,t)\right\vert _{x}\circ\chi\right)\cdot\left.\frac{\partial\chi}{\partial t}\right\vert _{X}.
\end{align*}
As a consequence, 
\[
\left(\left.\frac{\partial T_{\alpha}}{\partial t}\right\vert _{X}\right)^{\overrightarrow{(\beta)}}=\left(\left.\frac{\partial\left(T_{\alpha}^{\overrightarrow{(\beta)}}(x,t)\right)}{\partial t}\right\vert _{x}\right)+\left(\left.\nabla_{x}T_{\alpha}^{\overrightarrow{(\beta)}}(x,t)\right\vert _{x}\right)\cdot\left.\frac{\partial\chi}{\partial t}\right\vert _{X}^{\overrightarrow{(\beta)}}.
\]

\section{Basic kinematic formulas\label{app:Basic-Kinematical-Formulas}}

In this section some useful kinematic formulas are proven (for a complete
presentation of this subject see e.g. \cite{Lebedev}). They are the
basis of the procedure on which Hamilton-Piola postulation is founded.
However, because of they central role, they cannot be avoided in any
case: their use can be only postponed to subsequent steps, when different
postulations are attempted and indeed kinematic formulas of this type
are presented in any textbook of continuum mechanics. From now on,
the $\alpha$ domain will coincide with the Lagrangian set of coordinates
while $\beta$ domain will coincide with the Eulerian domain and the
notation $\left(\cdot\right)^{\overrightarrow{\left(\mathcal{B}\right)}}$
and $\left(\cdot\right)^{\overrightarrow{\left(\mathcal{E}\right)}}$
will be consistently used. They will be omitted occasionally for the
sake of readability.

\subsection{Formulas on Eulerian mass density and its gradients}

Mass density and its gradients play a pivotal role in the strain energy
of fluids. Here we gather some useful formulas relating them to $C\mbox{, }F$
and $\nabla F$ (we will omit the needed $\left(\cdot\right)^{\overrightarrow{\left(\mathcal{B}\right)}},\left(\cdot\right)^{\overrightarrow{(\mathcal{E})}})$
for the sake of king readability).

\subsubsection{The derivative of the determinant a matrix with respect its entries}

We start by recalling the well-known formula 
\[
\frac{\partial\det(A)}{\partial A_{M}^{i}}=\det A\left(A^{-T}\right)_{i}^{M},
\]
which can be recovered by using the Laplace rule for calculating the
determinant 
\[
\delta_{M}^{N}\det A=A_{M}^{a}\left(A^{\ast}\right)_{a}^{N},
\]
where $\left(A^{\ast}\right)_{i}^{N}$ is the cofactor of the element
$A_{N}^{i}$. Observing that the cofactors of all elements of the
$M-th$ row are independent of the entry $A_{M}^{i}$, together with
the inversion theorem for matrices, one gets 
\[
\frac{\partial\det(A)}{\partial A_{M}^{i}}=\left(A^{\ast}\right)_{i}^{M}=\det A\left(A^{-T}\right)_{i}^{M}.
\]

\subsubsection{Partial derivatives of $\rho,$ $J$ and $F^{-1}$ with respect to
$F$\label{app:derhodf}}

Once one recalls that 
\[
\rho_{0}\det F=\rho,
\]
and having defined the cofactor of $F$ as 
\[
\left(F^{\ast}\right)_{i}^{A}F_{A}^{j}=\det F\delta_{i}^{j}
\]
it is easy to deduce that 
\begin{eqnarray}
\frac{\partial J}{\partial F_{M}^{i}} & = & J\left(F^{-T}\right)_{M}^{i}=\frac{\rho_{0}}{\rho}\left(F^{-T}\right)_{M}^{i},\nonumber \\
\frac{\partial\rho}{\partial F_{M}^{i}} & = & -\rho\left(F^{-1}\right)_{i}^{M},\label{DERHODEF}\\
\frac{\partial\left(F^{-1}\right)_{j}^{N}}{\partial F_{M}^{i}} & = & -\left(F^{-1}\right)_{i}^{N}\left(F^{-1}\right)_{j}^{M}.\label{DEFMENO1DEF}
\end{eqnarray}

\subsubsection{Partial derivative of mass density with respect to $C$}

To prove the identity 
\begin{equation}
\frac{\partial\rho}{\partial C_{MN}}=-\frac{\rho}{2}\left(F^{-1}\right)^{Ma}\left(F^{-1}\right)_{a}^{N},\label{DERIVATIVEMASSGREEN}
\end{equation}
we proceed in the following way: 
\[
\frac{\partial\rho}{\partial C_{MN}}=\rho_{0}\frac{\partial\left(\det C\right)^{-\frac{1}{2}}}{\partial C_{MN}}=\rho_{0}\frac{\partial\left(\det C\right)^{-\frac{1}{2}}}{\partial\det C}\frac{\partial\det C}{\partial C_{MN}}=-\frac{\rho_{0}}{2}\left(\det C\right)^{-\frac{3}{2}}\frac{\partial\det C}{\partial C_{MN}}.
\]
In conclusion we have 
\[
\frac{\partial\rho}{\partial C_{MN}}=-\frac{\rho_{0}}{2}\left(\det C\right)^{-\frac{1}{2}}\left(C^{-1}\right)^{MN}=-\frac{\rho}{2}\left(F^{-1}\right)^{Ma}\left(F^{-1}\right)_{a}^{LN}.
\]

\subsubsection{Lagrangian and Eulerian gradients of $F^{-1}$}

Starting from 
\[
\left(F^{-1}\right)_{a}^{M}F_{N}^{a}=\delta_{N}^{M},
\]
after differentiation we obtain: 
\[
\left(F^{-1}\right)_{a}^{M}F_{N,O}^{a}+F_{N}^{a}\left(F^{-1}\right)_{a,O}^{M}=0,
\]
which produces the following chain of equalities: 
\begin{eqnarray}
F_{N}^{a}\left(F^{-1}\right)_{a,O}^{M} & = & -\left(F^{-1}\right)_{a}^{M}F_{N,O}^{a}\nonumber \\
\left(F^{-1}\right)_{i,O}^{M} & = & -\left(F^{-1}\right)_{i}^{A}\left(F^{-1}\right)_{a}^{M}F_{A,O}^{a}\label{GRADFMENO1LAGRANGIAN}
\end{eqnarray}
The last equality can be then multiplied by $F^{-1}$ to get the Eulerian
gradient 
\[
\left(F^{-1}\right)_{i,j}^{M}=-\left(F^{-1}\right)_{j}^{A}\left(F^{-1}\right)_{i}^{B}\left(F^{-1}\right)_{a}^{M}F_{B,A}^{a}.
\]
It can be useful to observe that: 
\begin{equation}
-\left(\rho\left(F^{-1}\right)_{i}^{M}\right)_{,j}=-\rho_{,j}\left(F^{-1}\right)_{i}^{M}-\rho\left(F^{-1}\right)_{i,j}^{M}=-\rho_{,j}\left(F^{-1}\right)_{i}^{M}+\rho\left(F^{-1}\right)_{j}^{A}\left(F^{-1}\right)_{i}^{B}\left(F^{-1}\right)_{a}^{M}F_{B,A}^{a}\label{EULGRADRHOFMENO1}
\end{equation}

\subsubsection{Expression of Eulerian gradient of density in terms of $F$ and its
gradients}

We start from the defining relationship: 
\begin{equation}
\rho=\frac{\rho_{0}}{\det\left(F\right)}=\rho_{0}\det\left(F^{-1}\right).\label{Density}
\end{equation}
As it is possible to assume that $\rho_{0}$ is constant, we calculate
the gradient of the density as follows 
\[
\rho_{,i}=\rho_{0}\det\left(F^{-1}\right)_{,i}=\rho_{0}\frac{\partial\det\left(F^{-1}\right)}{\partial\left(F^{-1}\right)_{a}^{A}}\frac{\partial\left(F^{-1}\right)_{a}^{A}}{\partial x^{i}}=\rho_{0}\det\left(F^{-1}\right)F_{A}^{a}\left(F^{-1}\right)_{a,i}^{A},
\]
and finally 
\[
\rho_{,i}^{\overrightarrow{\left(\mathcal{B}\right)}}=\rho F_{A}^{b}\left(F^{-1}\right)_{b,B}^{A}\left(F^{-1}\right)_{i}^{B}.
\]
To summarize, from all the previous expressions we obtain the following
useful formulas :

\begin{eqnarray}
\frac{\rho_{,i}}{\rho} & = & -\left(F^{-1}\right)_{a}^{A}\left(F^{-1}\right)_{i}^{B}F_{A,B}^{a}=-\left(F^{-1}\right)_{a}^{A}F_{A,i}^{a}\label{GRADRHOGRAF}\\
\rho_{,i} & = & \rho F_{A}^{a}\left(F^{-1}\right)_{i}^{B}\left(F^{-1}\right)_{a,B}^{A}=\rho F_{A}^{a}\left(F^{-1}\right)_{a,i}^{A}\nonumber \\
F_{A}^{a}\left(F^{-1}\right)_{a,M}^{A} & = & \frac{\rho_{,i}}{\rho}F_{M}^{i}\nonumber \\
\left(F^{-1}\right)_{j,A}^{M}\left(F^{-1}\right)_{i}^{A} & = & \left(F^{-1}\right)_{j}^{M}\frac{\rho_{,i}}{\rho}.\nonumber 
\end{eqnarray}

\subsubsection{Calculation of partial derivative of Eulerian gradient of mass density
with respect to $F$}

We need to estimate the following partial derivative: 
\[
\frac{\partial\rho_{,i}}{\partial F_{M}^{j}}=\frac{\partial}{\partial F_{M}^{j}}\left(-\rho\left(F^{-1}\right)_{a}^{A}\left(F^{-1}\right)_{i}^{B}\right)F_{A,B}^{a}.
\]
As we have that

\begin{gather*}
\left(\frac{\partial\rho}{\partial F_{M}^{i}}\left(F^{-1}\right)_{j}^{N}\left(F^{-1}\right)_{k}^{O}+\rho\left(F^{-1}\right)_{k}^{O}\frac{\partial\left(F^{-1}\right)_{j}^{N}}{\partial F_{M}^{i}}+\rho\left(F^{-1}\right)_{j}^{N}\frac{\partial\left(F^{-1}\right)_{k}^{O}}{\partial F_{M}^{i}}\right)=\\
-\rho\left(F^{-1}\right)_{i}^{M}\left(F^{-1}\right)_{j}^{N}\left(F^{-1}\right)_{k}^{O}-\rho\left(F^{-1}\right)_{i}^{N}\left(F^{-1}\right)_{j}^{M}\left(F^{-1}\right)_{k}^{O}-\rho\left(F^{-1}\right)_{j}^{N}\left(F^{-1}\right)_{i}^{O}\left(F^{-1}\right)_{k}^{M},
\end{gather*}
where we used the equalities \eqref{DERHODEF} and \eqref{DEFMENO1DEF},
we can then conclude 
\[
\frac{\partial\rho,_{i}}{\partial F_{M}^{j}}=\rho\left(\left(F^{-1}\right)_{j}^{M}\left(F^{-1}\right)_{i}^{A}\left(F^{-1}\right)_{a}^{B}F_{B,A}^{a}+\left(F^{-1}\right)_{j}^{C}\left(F^{-1}\right)_{i}^{M}\left(F^{-1}\right)_{b}^{D}F_{D,C}^{b}+\left(F^{-1}\right)_{i}^{E}\left(F^{-1}\right)_{j}^{F}\left(F^{-1}\right)_{c}^{M}F_{F,E}^{c}\right),
\]
by using \eqref{GRADRHOGRAF} we get 
\[
\frac{\partial\rho,_{i}}{\partial F_{M}^{j}}=-\rho_{,i}\left(F^{-1}\right)_{j}^{M}-\rho_{,j}\left(F^{-1}\right)_{i}^{M}+\rho\left(F^{-1}\right)_{i}^{A}\left(F^{-1}\right)_{lj}^{B}\left(F^{-1}\right)_{a}^{M}F_{B,A}^{a}.
\]
Finally by substituting \eqref{EULGRADRHOFMENO1} we can conclude:
\begin{equation}
\frac{\partial\rho,_{i}}{\partial F_{M}^{j}}=-\rho_{,j}\left(F^{-1}\right)_{i}^{M}-\left(\rho\left(F^{-1}\right)_{j}^{M}\right)_{,i}\label{DEGRADRHODEFFINAL}
\end{equation}

\subsubsection{The derivatives of $\left(\beta\right)^{\protect\overrightarrow{\left(\mathcal{B}\right)}}$
with respect $F$ and $\nabla F$\label{app:dbetadef}}

We start from a direct expression for $\left(\beta\right)^{\overrightarrow{\left(\mathcal{B}\right)}}$:
\[
\left(\beta\right)^{\overrightarrow{\left(\mathcal{B}\right)}}=\left(\nabla\rho\cdot\nabla\rho\right)^{\overrightarrow{\left(\mathcal{B}\right)}}=\left(g^{ab}\rho_{,a}\rho_{,b}\right)^{\overrightarrow{\left(\mathcal{B}\right)}},
\]
which implies that 
\begin{eqnarray}
\frac{\partial}{\partial F}\left(\beta\right)^{\overrightarrow{\left(\mathcal{B}\right)}} & = & 2\left(\nabla\rho\right)^{\overrightarrow{\left(\mathcal{B}\right)}}\cdot\frac{\partial\left(\nabla\rho\right)^{\overrightarrow{\left(\mathcal{B}\right)}}}{\partial F},\label{DEBETADEFINITIAL}\\
\frac{\partial}{\partial\nabla F}\left(\beta\right)^{\overrightarrow{\left(\mathcal{B}\right)}} & = & 2\left(\nabla\rho\right)^{\overrightarrow{\left(\mathcal{B}\right)}}\cdot\frac{\partial\left(\nabla\rho\right)^{\overrightarrow{\left(\mathcal{B}\right)}}}{\partial\nabla F}.\label{DEBETADEGRADFINITIAL}
\end{eqnarray}
Then using \eqref{DEBETADEF} and \eqref{DEGRADRHODEFFINAL} we get
easily: 
\begin{eqnarray}
\frac{\partial\beta}{\partial F_{M}^{i}} & = & 2g^{ab}\rho_{,a}\frac{\partial\left(\rho,_{b}\right)^{\overrightarrow{\left(\mathcal{B}\right)}}}{\partial F_{M}^{i}},\nonumber \\
\frac{\partial\left(\beta\right)^{\overrightarrow{\left(\mathcal{B}\right)}}}{\partial F_{M}^{i}} & = & -2g^{ab}\left(\rho_{,a}\rho_{,i}\left(F^{-1}\right)_{b}^{M}+\rho_{,a}\left(\rho\left(F^{-1}\right)_{i}^{M}\right)_{,b}\right)^{\overrightarrow{\left(\mathcal{B}\right)}}.\label{DEBETADEF}
\end{eqnarray}

Similarly, using \eqref{DEBETADEGRADFINITIAL} and \eqref{GRADRHOGRAF}
we obtain 
\begin{equation}
\frac{\partial\left(\beta\right)^{\overrightarrow{\left(\mathcal{B}\right)}}}{\partial F_{M,N}^{i}}=2g^{ab}\left(\rho_{,a}\right)^{\overrightarrow{\left(\mathcal{B}\right)}}\frac{\partial\left(\rho,_{b}\right)^{\overrightarrow{\left(\mathcal{B}\right)}}}{\partial F_{M,N}^{i}}=-2g^{ab}\left(\rho\rho_{,a}\left(F^{-1}\right)_{i}^{M}\left(F^{-1}\right)_{b}^{N}\right)^{\overrightarrow{\left(\mathcal{B}\right)}}.\label{DEBETADEGRADF}
\end{equation}

\subsection{Derivatives of $C,C^{-1},\nabla C$ and $\nabla C^{-1}$ with respect
to $F$ and $\nabla F$}

\subsubsection{Computation of\textmd{ $\frac{\partial C_{MN}}{\partial F_{P}^{i}}$}
\label{sub:DCDF}}

\begin{align*}
\frac{\partial C_{MN}}{\partial F_{P}^{i}} & =g_{ab}\frac{\partial}{\partial F_{P}^{i}}\left(F_{M}^{a}F_{N}^{b}\right)=g_{ab}\left(\frac{\partial F_{M}^{b}}{\partial F_{P}^{i}}F_{N}^{a}+F_{M}^{b}\frac{\partial F_{N}^{a}}{\partial F_{P}^{i}}\right)\\
 & =g_{ab}\left(\delta_{i}^{b}\delta_{M}^{P}F_{N}^{a}+F_{M}^{b}\delta_{i}^{a}\delta_{N}^{P}\right)=\left(\delta_{M}^{P}F_{iN}+F_{iM}\delta_{N}^{P}\right).
\end{align*}

\subsubsection{Computation of \textmd{$\frac{\partial C_{MN,O}}{\partial F_{P}^{i}}$}}

\begin{eqnarray*}
\frac{\partial C_{MN,O}}{\partial F_{P}^{i}} & = & \frac{\partial}{\partial F_{P}^{i}}\left(\frac{\partial F_{M}^{a}}{\partial X^{O}}F_{Na}+\frac{\partial F_{N}^{b}}{\partial X^{O}}F_{Lb}\right)\\
 & = & \frac{\partial}{\partial F_{P}^{i}}\left(\frac{\partial F_{M}^{a}}{\partial X^{O}}\right)F_{Na}+\frac{\partial F_{M}^{b}}{\partial X^{O}}\frac{\partial F_{Nb}}{\partial F_{P}^{i}}+\frac{\partial}{\partial F_{P}^{i}}\left(\frac{\partial F_{N}^{c}}{\partial X^{O}}\right)F_{Mc}+\frac{\partial F_{N}^{d}}{\partial X^{O}}\frac{\partial F_{Ml}}{\partial F_{P}^{d}}\\
 & = & g_{ab}F_{M,O}^{a}\frac{\partial F_{N}^{b}}{\partial F_{P}^{i}}+g_{cd}F_{N,O}^{c}\frac{\partial F_{M}^{d}}{\partial F_{P}^{i}}=g_{ab}F_{M,O}^{a}\delta_{i}^{b}\delta_{P}^{N}+g_{cd}F_{N,O}^{c}\delta_{i}^{d}\delta_{P}^{M}\\
 & = & F_{iM,O}\delta_{P}^{N}+F_{iN,O}\delta_{P}^{M}.
\end{eqnarray*}

\subsubsection{Computation of $\frac{\partial C_{MN}^{-1}}{\partial F_{P}^{i}}$}

\begin{eqnarray*}
\frac{\partial\left(C^{-1}\right)_{MN}}{\partial F_{P}^{i}} & = & \frac{\partial\left(\left(F^{-1}\right)_{aM}\left(F^{-1}\right)_{N}^{a}\right)}{\partial F_{P}^{i}}=\frac{\partial\left(\left(F^{-1}\right)_{aM}\right)}{\partial F_{P}^{i}}\left(F^{-1}\right)_{N}^{a}+\left(F^{-1}\right)_{bM}\frac{\partial\left(\left(F^{-1}\right)_{N}^{b}\right)}{\partial F_{P}^{i}}.
\end{eqnarray*}
Using equation \eqref{DEFMENO1DEF} we obtain

\begin{eqnarray*}
\frac{\partial\left(C^{-1}\right)_{MN}}{\partial F_{P}^{i}} & = & -\left(F^{-1}\right)_{Mi}\left(F^{-1}\right)_{a}^{P}\left(F^{-1}\right)_{N}^{a}-\left(F^{-1}\right)_{Ni}\left(F^{-1}\right)^{bP}\left(F^{-1}\right)_{bM}.
\end{eqnarray*}

\subsubsection{Computation of $\frac{\partial C_{MN,O}^{-1}}{\partial F_{P}^{i}}$ }

\begin{eqnarray*}
\frac{\partial C_{MN,O}^{-1}}{\partial F_{P}^{i}} & = & \left(\left(F^{-1}\right)_{Ma,O}\frac{\partial\left(F^{-1}\right)_{N}^{a}}{\partial F_{P}^{i}}+\frac{\partial\left(F^{-1}\right)_{Mb}}{\partial F_{P}^{i}}\left(F^{-1}\right)_{N,O}^{b}\right)\\
 & = & -\left(\left(F^{-1}\right)_{Ni}\left(F^{-1}\right)^{aP}\left(F^{-1}\right)_{Ma,O}+\left(F^{-1}\right)_{Mi}\left(F^{-1}\right)^{aP}\left(F^{-1}\right)_{aN,O}\right)\\
 & = & -\left(F^{-1}\right)^{aP}\left(\left(F^{-1}\right)_{Ni}\left(F^{-1}\right)_{Ma,O}+\left(F^{-1}\right)_{Mi}\left(F^{-1}\right)_{aN,O}\right).
\end{eqnarray*}

\subsubsection{Computation of $\frac{\partial C_{MN,O}}{\partial F_{P,Q}^{i}}$}

The computation is straightforward

\[
\frac{\partial C_{MN,O}}{\partial F_{P,Q}^{i}}=\frac{\partial}{\partial F_{P,Q}^{i}}\left(F_{M,O}^{a}F_{Na}+F_{N,O}^{b}F_{Mb}\right)=\left(\delta_{i}^{a}\delta_{M}^{P}\delta_{O}^{Q}F_{Na}+\delta_{i}^{b}\delta_{N}^{P}\delta_{O}^{Q}F_{Mb}\right)=\left(\delta_{M}^{P}\delta_{O}^{Q}F_{Ni}+\delta_{N}^{P}\delta_{O}^{Q}F_{Mi}\right).
\]

\subsubsection{Computation of\textmd{ $\frac{\partial C_{MN,O}^{-1}}{\partial F_{P,Q}^{i}}$}}

We compute the partial derivative as the following product: 
\[
\frac{\partial C_{MN,O}^{-1}}{\partial F_{P,Q}^{i}}=\frac{\partial C_{MN,O}^{-1}}{\partial\left(F^{-1}\right)_{A,B}^{a}}\frac{\partial\left(F^{-1}\right)_{A,B}^{a}}{\partial F_{P,Q}^{i}}.
\]
The first term is directly proceed:

\begin{eqnarray*}
\frac{\partial C_{MN,O}^{-1}}{\partial\left(F^{-1}\right)_{P,Q}^{i}} & = & \frac{\partial}{\partial\left(F^{-1}\right)_{P,Q}^{i}}\left(g_{ab}\left(F^{-1}\right)_{N}^{a}\left(F^{-1}\right)_{M,O}^{b}+\left(F^{-1}\right)_{Mc}\left(F^{-1}\right)_{N,O}^{c}\right)\\
 & = & g_{ab}\delta_{i}^{b}\delta_{M}^{P}\delta_{Q}^{O}\left(F^{-1}\right)_{N}^{a}+\delta_{i}^{c}\delta_{N}^{P}\delta_{Q}^{O}\left(F^{-1}\right)_{Mc}\left(F^{-1}\right)_{N,L}^{c}\\
 & = & \delta_{Q}^{O}\left[\delta_{M}^{P}\left(F^{-1}\right)_{iN}+\delta_{N}^{P}\left(F^{-1}\right)_{Mi}\right].
\end{eqnarray*}
Deriving equation \eqref{GRADFMENO1LAGRANGIAN} with respect to $F_{P,Q}^{i}$we
obtain

\[
\frac{\partial\left(F^{-1}\right)_{i,N}^{M}}{\partial F_{P,Q}^{j}}=-\left(F^{-1}\right)_{j}^{M}\left(F^{-1}\right)_{i}^{P}\delta_{N}^{Q},
\]
Combining the results and considering that

\[
\left(F^{-1}\right)_{M,N}^{i}=g^{ia}g_{MA}\left(F^{-1}\right)_{a,N}^{A},
\]
we finally have 
\begin{align*}
\frac{\partial C_{MN,O}^{-1}}{\partial F_{P,Q}^{i}} & =-\delta_{A}^{O}\left[\delta_{M}^{B}\left(F^{-1}\right)_{aN}+\delta_{N}^{B}\left(F^{-1}\right)_{Ma}\right]\left(F^{-1}\right)_{Bi}\left(F^{-1}\right)^{aP}\delta_{Q}^{A}\\
 & =-\delta_{Q}^{O}\left[\left(F^{-1}\right)_{Mi}\left(F^{-1}\right)^{aP}\left(F^{-1}\right)_{aN}+\left(F^{-1}\right)_{Ni}\left(F^{-1}\right)^{bP}\left(F^{-1}\right)_{Mb}\right].
\end{align*}

\section{Gauss divergence theorem for embedded Riemannian manifolds}

We choose a global orthonormal basis $\left(e_{i},i=1,2,3\right)$
for the vector field of displacements in $E^{3}$, the tridimensional
Euclidean space. All tensor fields will be represented by their components
with respect to this basis. In this section we consider an embedded
Riemannian manifold $\mathcal{M}$ in $E^{3}$. This manifold can
be therefore a regular curve or surface, but will be restricted to
a surface in the present discussion. As $\mathcal{M}$ can be equipped
with a Gaussian coordinate systems, it is possible to introduce in
the neighborhood of any point of $\mathcal{M}$ (For more details
see dell'Isola et al.\ \cite{dellisolaseppechermadeo2012}): 
\begin{itemize}
\item $P$, the field of projection operator on tangent space; 
\item $Q$ the field of projection operator on tangent space. 
\end{itemize}
These projectors verify the following obvious identities:

\begin{gather*}
\delta_{i}^{j}=P_{i}^{j}+Q_{i}^{j},\quad P_{i}^{a}P_{a}^{j}=P_{i}^{j},\\
Q_{i}^{a}Q_{a}^{j}=Q_{i}^{j},\quad P_{i}^{a}Q_{a}^{j}=0.
\end{gather*}

In order to simplify the forthcoming calculations, instead of using
curvilinear coordinates, we rather use a global Cartesian coordinate
system, completed by $P$ and $Q$ in the neighborhood of $\mathcal{M}$.
This technical choice is exactly the same one which allowed Germain
to generalize, for second gradient materials, the results found by
Green, Rivlin, Toupin and Mindlin.

The unit external normal to $\mathcal{M}$ on its border, which is
denoted $\nu$, belongs to the tangent space to $\mathcal{M}$.

Using these notations the divergence theorem reads (see e.g. Spivak
\cite{Spivak}): for any vector field $W$ defined in the vicinity
of $M$ 
\begin{equation}
\int_{M}(P_{b}^{a}W^{b})_{,c}P_{a}^{c}dS=\int_{\partial M}W^{a}P_{a}^{b}\nu_{b}dL\label{Divergence}
\end{equation}
This theorem together with relation 
\[
Q_{j,b}^{a}P_{a}^{b}=-Q_{j}^{a}P_{a,b}^{b}
\]
implies that, for any vector field $W$ defined in a neighborhood
of $\mathcal{M}$, 
\[
\begin{aligned}\int_{M}\left(W^{a}\right),_{b}P_{a}^{b}dS & =\int_{M}\left[(P_{b}^{a}W^{b})_{,c}P_{a}^{c}+(Q_{e}^{d}W^{e})_{,f}P_{d}^{f}\right]dS\\
 & =\int_{M}W^{a}Q_{a,c}^{b}P_{b}^{c}dS+\int_{\partial M}W^{d}P_{d}^{e}\nu_{e}dL=-\int_{M}W^{a}Q_{a}^{b}P_{b,c}^{c}dS+\int_{\partial M}W^{d}P_{d}^{f}\nu_{f}dL.
\end{aligned}
\]

\end{document}